\documentclass[11pt,a4paper]{article}
\pdfoutput=1
\usepackage{jheppub}
\usepackage{comment}
\usepackage{gensymb}
\usepackage{longtable}
\usepackage{amssymb}
\usepackage{lipsum}
\DeclareSymbolFont{matha}{OML}{txmi}{m}{it}% txfonts
\DeclareMathSymbol{\varv}{\mathord}{matha}{118}
\usepackage{amsmath}
\usepackage{amssymb}
\usepackage{graphicx}
\usepackage{subfigure}
\usepackage{pstricks}
\usepackage{bm}
\usepackage{lscape}
\usepackage{pbox}
\usepackage{placeins}
\usepackage{booktabs}
\usepackage[T1]{fontenc}
\usepackage{footnote}
\usepackage{pdfpages}
\usepackage{hhline}
\usepackage{multirow}
\usepackage{multicol}
\usepackage{enumitem}
\usepackage[toc,page]{appendix}
\allowdisplaybreaks
\usepackage{comment}
\usepackage{float}                                              
\usepackage{soul}
\setstcolor{red}
\usepackage{textcomp}
\usepackage{gensymb}
\usepackage{xcolor,pifont}
\newcommand{\xmark}{\text{\ding{55}}}
\newcommand{\cmark}{\text{\ding{51}}}
\newcommand*\colourcheck[1]{%
  \expandafter\newcommand\csname #1check\endcsname{\textcolor{#1}{\ding{52}}}%
}
\colourcheck{blue}
\colourcheck{green}
\colourcheck{red}
\colourcheck{olivegreen}

\usepackage{empheq}
\usepackage[numbers]{natbib}
\usepackage{notoccite}

\usepackage{rotating}
\usepackage{tabularx}
\usepackage[labelfont=bf]{caption} % optional

\FloatBarrier
\renewcommand{\thetable}{\arabic{table}}

\newcommand{\beq}{\begin {equation}}  
\newcommand{\eeq}{\end   {equation}} 
\newcommand{\bea}{\begin {eqnarray}} 
\newcommand{\eea}{\end   {eqnarray}}  
\newcommand{\baa}{\begin {array}   } 
\newcommand{\eaa}{\end   {array}   }     
\newcommand{\bit}{\begin {itemize} }
\newcommand{\eit}{\end   {itemize} }
\newcommand{\be }{\begin {equation}} 
\newcommand{\ee }{\end   {equation}}

\definecolor{MyDarkBlue}{rgb}{0.1, 0.1, 0.8} %defining the color 'MyDarkBlue'
\definecolor{SBlue}{rgb}{0.2, 0.4, 0.7} %defining the color 'MyDarkBlue'
\definecolor{MyLightBlue}{rgb}{0.22,0.51,0.9}
\definecolor{MyGreen}{rgb}{0.0, 0.5, 0.3}
\definecolor{BrickRed}{rgb}{0.8, 0.25, 0.33}
\RequirePackage{hyperref}
\hypersetup{colorlinks, citecolor=BrickRed,linkcolor=MyDarkBlue, urlcolor=MyLightBlue}

\begin{document}
%\vspace*{-0.2in}
\preprint{OSU-HEP-19-09}
%\color{SBlue}{\bf OSU-HEP-19-04}
%\end{flushright}
%\vspace{0.5cm}
%\begin{center}
%{\Large\bf 
\title{\textbf{Minimal Realizations of  Dirac Neutrino Mass from Generic One-loop and Two-loop Topologies at $d=5$} 
 }
 \author[a]{Sudip Jana,}
 \author[b]{Vishnu P.K.,}
 \author[b]{Shaikh Saad}
 
 \affiliation[a]{Max-Planck-Institut f{\"u}r Kernphysik, Saupfercheckweg 1, 69117 Heidelberg, Germany}
 \affiliation[b]{Department of Physics, Oklahoma State University, Stillwater, OK 74078, USA}

\emailAdd{ sudip.jana@mpi-hd.mpg.de,  vipadma@okstate.edu, shaikh.saad@okstate.edu}
%%%%%%%%%%%%%%%%%%%%%%%%%%%%%%%%%%%%%%%%%%%%%%%
%%%%%%%%%%%%%%%%%%%%%%%%%%%%%%%%%%%%%%%%%%%%%%%
\abstract{
We carry out a systematic investigation for the minimal  Dirac neutrino mass models emerging from generic one-loop and two-loop topologies  that arise from $d=5$ effective operator  with a singlet scalar, $\sigma$. To ensure that the tree-level Dirac mass, as well as Majorana mass terms at all orders, are absent for the neutrinos, we work in the framework  where the Standard Model is supplemented by the well-motivated $U(1)_{B-L}$ gauge symmetry.
At the one-loop level, we analyze six possible topologies, out of which two of them have the potential to generate desired Dirac neutrino mass. Adopting a systematic approach to select  minimal models, we construct seventeen viable one-loop Dirac neutrino mass models. By embracing a similar methodical approach at the two-loop, we work out twenty-three  minimal candidates. Among the forty selected economical models, the  majority of the  models proposed in this work are new.  In our search, we  also include the scenarios where the particles in the loop carry charges under the color group. Furthermore, we discuss the possible dark matter candidates within a given model, if any, without extending the minimal particle content. }

\maketitle
%\newpage
%{\hypersetup{linkcolor=black}\tableofcontents}
\setcounter{footnote}{0}

%%%%%%%%%%%%%%%%%%%%%%%%%%%%%%%%%%%%%%%%%%%%%%%
%%%%%%%%%%%%%%%%%%%%%%%%%%%%%%%%%%%%%%%%%%%%%%%
\section{Introduction}\label{SEC-01}
The Standard Model (SM) of particle physics has been a very successful theory to explain the observed universe at the microscopic level. Despite its great triumph, the SM is unable to explain several observed phenomena, among them, the observation of neutrino oscillations is considered to be the biggest drawback of the SM. Neutrinos are still the least understood of all the fundamental particles discovered so far.  The origin of the neutrino oscillations must be linked to new physics beyond the SM (BSM), hence finding the  mechanism responsible for neutrino mass generation is of greatest importance.   However, to attempt to find a mechanism behind the neutrino mass, the first obstacle  the theorists encounter is the unknown nature of the neutrinos. In principle, neutrinos can be Majorana or Dirac type in nature, with no theoretical preference toward either of the possibilities. Most of the proposals in the literature assume that neutrinos are Majorana\footnote{Implementation of seesaw mechanisms for generating Majorana neutrino masses are proposed in Refs. \cite{Minkowski:1977sc, Yanagida:1979as, GellMann:1980vs, Mohapatra:1979ia, Schechter:1980gr, Schechter:1981cv, Foot:1988aq} and radiative mass models for Majorana neutrinos are introduced in Refs.  \cite{Zee:1980ai, Zee:1985id, Babu:1988ki, Babu:1988ig, Ma:2006km}. 
For a recent review on Majorana neutrino mass models see Ref. \cite{Cai:2017jrq}.} in nature, whereas mass generation mechanisms for Dirac\footnote{The first radiative neutrino mass model was proposed in Ref. \cite{Cheng:1977ir}, assuming Dirac nature for the  neutrinos.} type neutrinos are less studied and deserve further attention. In the last several years, there have been growing interests\footnote{For construction of  Dirac neutrino mass models by employing seesaw mechanisms see for example Refs.  \cite{Ma:2014qra, Ma:2015mjd, Ma:2015raa, Valle:2016kyz, Chulia:2016ngi, Chulia:2016giq, Reig:2016ewy, CentellesChulia:2017koy, CentellesChulia:2017sgj, Borah:2017leo, Bonilla:2017ekt, Borah:2017dmk, Borah:2018nvu, Ma:2018bow,  Borah:2019bdi, Gu:2019ogb, Calle:2019mxn}. For works on radiative Dirac neutrino mass models see for example Refs. \cite{Mohapatra:1987hh, Mohapatra:1987nx, Balakrishna:1988bn, Branco:1978bz, Babu:1988yq, Gu:2007ug, Farzan:2012sa, Okada:2014vla, Bonilla:2016diq, Wang:2016lve, Ma:2017kgb, Wang:2017mcy, Helo:2018bgb, Reig:2018mdk, Han:2018zcn, Kang:2018lyy, Bonilla:2018ynb, Calle:2018ovc, Carvajal:2018ohk, Ma:2019yfo, Bolton:2019bou, Saad:2019bqf, Bonilla:2019hfb, Dasgupta:2019rmf, Jana:2019mez, Enomoto:2019mzl, Ma:2019byo,Restrepo:2019soi}.} in model building, assuming Dirac nature of the neutrinos.   

If neutrinos are Dirac particles, the theory must contain the right-handed neutrinos $\nu_R$, which are singlets under the SM. As a result, one can write down the following $d=4$ renormalizable term in the Lagrangian:
\begin{align}
\mathcal{L}_4= -y^{\nu}_{ij}\;\overline{L}_i \widetilde{H}\;{\nu_R}_j +h.c., \label{d4}
\end{align}
here, $L=(\nu_L\; \ell_L^-)^T$ is the SM lepton doublet, $H=(H^+\; H^0)^T$ is the SM Higgs doublet, $\widetilde{H}=\epsilon H^{\ast}$ ($\epsilon$ is the 2-index Levicivita tensor) and $i,j=1-3$ correspond to generation indices. Note that when the electroweak (EW) symmetry is spontaneously broken by the vacuum expectation value (VEV) of $H$, the $d=4$ term given in Eq. \eqref{d4} gives Dirac mass to the neutrinos at the tree-level. However, from the neutrino oscillation data, this demands that the associated Yukawa couplings must be extremely small, $y^{\nu}\sim \mathcal{O} (10^{-11})$. Due to this small Yukawa couplings, this possibility  is expected not to be natural \cite{tHooft:1979rat}.  Instead, it is aesthetically attractive, to generate Dirac mass for the neutrinos radiatively that allows natural values of the Yukawa coupling as a result of the loop suppression. Implementation of such a mechanism requires additional symmetries BSM to forbid the tree-level mass term given in Eq. \eqref{d4}, as well as Majorana neutrino mass terms at all orders.   In principle, this additional symmetry can be discrete or continuous, global or local in nature. In the literature, many different symmetries \cite{Mohapatra:1987hh, Mohapatra:1987nx, Balakrishna:1988bn, Branco:1978bz, Babu:1988yq, Gu:2007ug, Farzan:2012sa, Okada:2014vla, Bonilla:2016diq, Wang:2016lve, Ma:2017kgb, Wang:2017mcy, Helo:2018bgb, Reig:2018mdk, Han:2018zcn, Kang:2018lyy, Bonilla:2018ynb, Calle:2018ovc, Carvajal:2018ohk, Ma:2019yfo, Bolton:2019bou, Saad:2019bqf, Bonilla:2019hfb, Dasgupta:2019rmf, Jana:2019mez, Enomoto:2019mzl, Ma:2019byo,Restrepo:2019soi} are considered to forbid these unwanted mass terms. 

Though many of the works assume global symmetries to prohibit    undesirable terms, from the theoretical grounds, introducing a local symmetry is more appealing. It is because the local symmetries are known to be respected by gravitational interactions \cite{Giddings:1987cg, Giddings:1989bq, Giddings:1988wv, Abbott:1989jw, Coleman:1989zu}, whereas global symmetries are not. Motivated by this feature of local symmetries, in this work we extend the SM with gauged $U(1)_{B-L}$ symmetry \cite{Davidson:1978pm, Marshak:1979fm, Mohapatra:1980qe, Wetterich:1981bx}.  Here $B$ and $L$ correspond to baryon and lepton numbers respectively. The introduced $U(1)_{B-L}$ symmetry not only serves the purpose of forbidding the aforementioned unwanted terms, but also it offers a rich phenomenology and is  considered as one of the most economical gauge extensions of the SM. However, in this work we do not focus on the associated phenomenology that are well studied in the literature\footnote{For a review see for example Ref.  \cite{Langacker:2008yv}.},  rather focus on neutrino mass generation mechanisms utilizing $U(1)_{B-L}$ symmetry.  It should be pointed out that the results of this work remain unchanged regardless of the nature (global or local) of this  symmetry.

It is well known that $U(1)_{B-L}$  gauge extension of the SM with three right-handed neutrinos is anomaly free.  Between two possible anomaly free $B-L$ charge assignments \cite{Montero:2007cd, Machado:2010ui, Machado:2013oza} of the right-handed neutrinos: $\nu_{R_{1,2,3}}=\{-1,-1,-1\}$ and $\nu_{R_{1,2,3}}=\{5,-4,-4\}$, only the latter is able to serve our purpose. Hence we adopt the second solution in this work. Consequently, the $d=4$ term of Eq. \eqref{d4} is forbidden, and we consider the following  $d=5$  effective  operator for neutrino mass generation:      
\begin{align}
\mathcal{L}_5= -\frac{h_{ij}}{\Lambda}\;\overline{L}_i \widetilde{H}\;{\nu_R}_j\sigma+h.c., \label{d5}
\end{align}
here, the neutral scalar $\sigma$ is a  singlet under the  SM and we assign it three units of $B-L$ charge, as a result, $h_{i1}=0$.  The quantum numbers of all the SM particles, the right-handed neutrinos, and the scalar singlet $\sigma$ under SM$\times U(1)_{B-L}$ are listed in Table \ref{charge}.

%%%%%%%%%%%%%%%%%%%%%%%%%%%%%%%%%%%%%%%%%%%%%%%%%%%%%%
\FloatBarrier
\begin{table}[t!]
\centering
\footnotesize
\resizebox{0.5\textwidth}{!}{
\begin{tabular}{|c|c|}
\hline
Multiplets& $SU(3)_C\times SU(2)_L\times U(1)_Y\times U(1)_{B-L}$   \\ \hline\hline
Quarks&
\pbox{10cm}{
\vspace{2pt}
${Q_L}_i (3,2,\frac{1}{6},\frac{1}{3})$\\
${u_R}_i (3,1,\frac{2}{3},\frac{1}{3})$\\
${d_R}_i (3,1,-\frac{1}{3},\frac{1}{3})$
\vspace{2pt}}
\\ \hline\hline
Leptons&
\pbox{10cm}{
\vspace{2pt}
${L}_i (1,2,-\frac{1}{2},-1)$\\
${\ell_R}_i (1,1,-1,-1)$\\
${\nu_R}_i (1,1,0,\{5,-4,-4\})$
\vspace{2pt}}
\\ \hline\hline
Higgs &
\pbox{10cm}{
\vspace{2pt}
$H (1,2,\frac{1}{2},0)$\\
$\sigma (1,1,0,3)$
\vspace{2pt}}
\\ \hline
\end{tabular}
}
\caption{ 
Quantum numbers of the SM particles, the right-handed neutrinos and the singlet scalar $\sigma$.
}\label{charge}
\end{table}
%%%%%%%%%%%%%%%%%%%%%%%%%%%%%%%%%%%%%%%%%%%%%%%%%%%%%

In this work, we aim to systematically search for the minimal models to generate Dirac neutrino mass radiatively by utilizing the $d=5$ effective operator given in Eq. \eqref{d5}. We construct generic topologies emerging from this $d=5$ operator  at the one-loop and two-loop levels and, build the associated minimal models. Whereas, only a few of these models resulting from our comprehensive systematic construction exist in the literature, however, most of the models presented in this work are new. Minimal one-loop and two-loop models resulting from our exhaustive search to be discussed at length  are summarized in Tables \ref{table-one}, \ref{table-one-color}, \ref{table-two} and  \ref{table-two-color}. Whereas a systematic classification of Majorana neutrino mass mechanisms and identifying minimal models are extensively discussed in the literature \cite{Ma:1998dn, Babu:2001ex, Bonnet:2009ej, Bonnet:2012kz, Law:2013dya, Sierra:2014rxa, Cepedello:2017eqf, Anamiati:2018cuq, Cepedello:2018rfh,  Klein:2019iws}, similar analysis for the case of  Dirac neutrinos are still lacking. Along this direction, some recent   attempts are taken: a systematic classification of tree-level and one-loop  mass mechanisms at $d=4$  \cite{Ma:2016mwh}, $d=5$ \cite{Yao:2018ekp, CentellesChulia:2018gwr},  
 $d=6$ \cite{Yao:2017vtm, CentellesChulia:2018bkz}, two-loop mass models at $d=4$ \cite{Ma:2016mwh, CentellesChulia:2019xky}, selecting minimal models at one-loop \cite{Calle:2018ovc} and identifying simplest models at one-loop, two-loop and three-loop \cite{Saad:2019bqf} are considered.

In our search strategy, \textit{minimality} refers to finding a model among many different possibilities such that it has the following features:
\begin{itemize}
\item It consists of minimum number of BSM states.

\item $SU(2)_L$ singlet BSM states are preferred. If BSM particles are required not to be iso-singlets, then we minimize the number of states that are charged under $SU(2)_L$.

\item If a particle is charged under $SU(2)_L$, then the lowest dimensional representation is preferred. We impose the same rule in scenarios, where  a particle carry $SU(3)_C$ charge. 

\item If possible, introduction of any BSM fermion is prohibited. 

\item For a model, where the presence of  BSM fermionic state is required, we assume it to be vector-like under SM$\times U(1)_{B-L}$ so that it does not alter the aforementioned anomaly-free conditions. 
\end{itemize}

\noindent
Moreover, whenever fermionic extension is required,  we always assume it comes with three generations and Dirac type in nature. When building a one-loop model, for the neutrinos it must not contain any tree-level mass term, and similarly, for a two-loop model, both tree-level and one-loop diagrams must be absent. A model can only be called a \textit{true model},  once these criteria are met.
Here we point out that, in this set-up, due to non-universal charge assignment of the right-handed neutrinos, one of the neutrinos will always remain massless, which however is completely consistent with neutrino oscillation data. Whereas by further extension of each of the models can accommodate non-zero  mass for all the neutrino states, and can be done straightforwardly, for the sake of minimality we do not discuss such possibilities here.  

In identifying minimal models emerging from generic one-loop and two-loop topologies, we first construct models by employing BSM fields that do not carry any color. Then we extend our search involving colored particles as well. By explicit construction, we demonstrate that for some of the models with colored particles in the loop, an analogous version with color-singlet states does not exist, or vice versa. For the  desired neutrino mass mechanism to realize,  we employ scalar leptoquarks, di-quarks and colored fermions depending on the specific model. Some of these colored particles  appear in various BSM theories such as grand unified theories\footnote{For the origin of neutrino mass from simple GUTs, see for example Refs.  \cite{Dorsner:2017wwn, Saad:2019vjo, Klein:2019jgb}.} (GUTs) \cite{Pati:1973uk, Pati:1974yy, Georgi:1974sy, Georgi:1974yf, Georgi:1974my, Fritzsch:1974nn}, technicolor models \cite{Dimopoulos:1979es,Dimopoulos:1979sp, Farhi:1980xs}, other compositeness scenarios \cite{Schrempp:1984nj},  R-parity violating supersymmetric models \cite{Barbier:2004ez},  and dark matter models \cite{Abercrombie:2015wmb}. 
Utilization of the colored particles in building Majorana neutrino masses are extensively considered by extending the particle content of the SM, see for example Refs.  \cite{Cai:2017jrq, Babu:2019mfe}. Colored particles have also received further attention in view of the possibilities  to explain certain striking discrepancies observed in the flavor sector. Models with scalar leptoquarks  can explain the discrepancies observed
mostly in rare decay modes of $B$ mesons by various experimental collaborations, like Belle, LHCb  and BaBar.  However, there is no literature where  colored scalars and/or fermions are introduced in the context of Dirac neutrino mass generation mechanism. For the first time, we are presenting radiative models where  colored particles, especially leptoquarks  are employed to generate tiny Dirac neutrino masses. The discovery of the leptoquarks would be an unambiguous signal of this kind of BSM physics and hence, various searches for such particles were conducted in the past experiments and the hunt
is still ongoing at the current collider experiments. 

Furthermore, for all minimal models proposed in this work, we investigate  possibilities of the existence of dark matter (DM) candidates within the working framework,  without extending the minimal particle content. Our analysis shows that  for majority of the proposed models,  the presence of the DM particles can arise naturally due to the appearance of residual dark symmetry resulting from spontaneously broken $U(1)_{B-L}$ group. Hence, these models can explain two seemingly uncorrelated  phenomena, the origin of neutrino mass and observed DM in the universe, under the same umbrella.  

The paper is organized as follows. In Section \ref{SEC-03}, we scrutinize the possible one-loop topologies and construct the minimal  Dirac neutrino mass models emerging from them. In Section \ref{SEC-04}, we investigate the two-loop possibilities and build the associated economical Dirac neutrino mass models from the generic topologies. We also investigate the possible dark matter candidates within the minimal one-loop and two-loop models in section \ref{SEC-dark}. Finally we conclude in Section \ref{SEC-con}.

%%%%%%%%%%%%%%%%%%%%%%%%%%%%%%%%%%%%%%%%%%%%%%%
%%%%%%%%%%%%%%%%%%%%%%%%%%%%%%%%%%%%%%%%%%%%%%%
\section{Search for minimal one-loop models}\label{SEC-03}
In this section, we systematically build minimal one-loop models in the framework introduced in Sec. \ref{SEC-01}. First,  we discuss the procedure of selecting relevant topologies from which minimal models can be constructed.
%%%%%%%%%%%%%%%%%%%%%%%%%%%%%%%%%%%%%%%%%%%%%%%
%%%%%%%%%%%%%%%%%%%%%%%%%%%%%%%%%%%%%%%%%%%%%%%
\subsection{Generic topologies, nomenclature and adopting viable topologies}
 Let us first define our convention. In this work, we denote each of the one-loop topologies by T1-x, where 1 unambiguously stands for one-loop  and x (x$=i, ii, iii, ...$)  is used to differentiate among the different possible  topologies. From each of these topologies, specifying the Lorentz structure can lead to multiple diagrams. If there exist more than one diagram within a fixed topology T1-x, we  label them as T1-x-y (with y= 1, 2, 3, ...) and furthermore, associated with a diagram   T1-x-y, since multiple models can be fabricated by varying the quantum numbers of the internal particles, we name model diagrams as T1-x-y-z (where z=A, B, C, ...). That is, a \textit{topology}  T1-x does not contain the Lorentz nature, whereas a \textit{diagram}, T1-x-y is formed from a \textit{topology} by specifying the Lorentz nature of the propagators. Furthermore, from a \textit{diagram} T1-x-y, we construct specific models by assigning  quantum numbers of each of the particles within the \textit{diagram} and call them \textit{model diagrams}  T1-x-y-z.  Whereas, the number of \textit{topologies} and \textit{diagrams} are finite, but infinite number of \textit{model diagrams} can be generated starting from a fixed \textit{topology}. However, we only focus on identifying the most economical \textit{model diagrams} emerging from each individual \textit{topology}.

\FloatBarrier
\begin{figure}[th!]
\centering
$$
\includegraphics[scale=0.35]{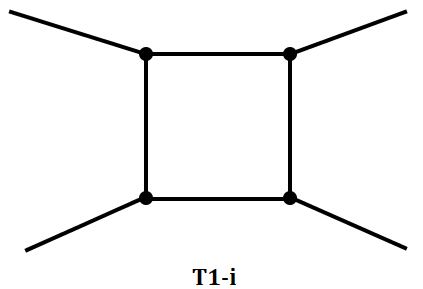}\hspace{0.12in}
\includegraphics[scale=0.35]{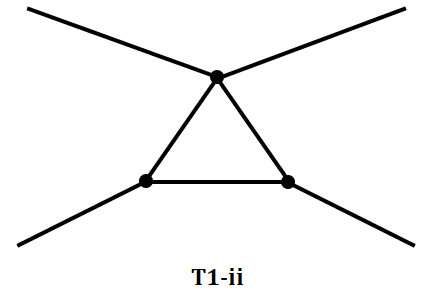}\hspace{0.12in}
\includegraphics[scale=0.35]{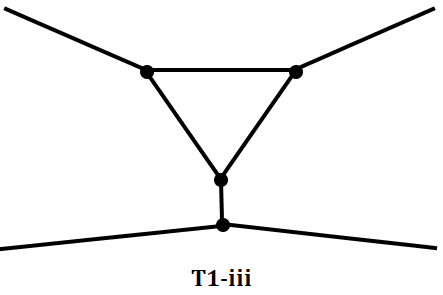}
$$
$$
\includegraphics[scale=0.35]{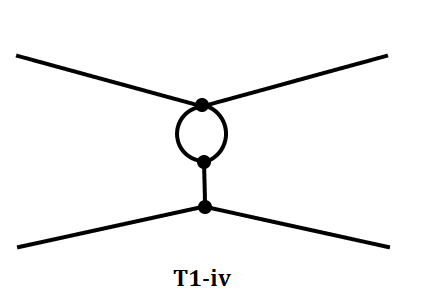}\hspace{0.12in}
\includegraphics[scale=0.35]{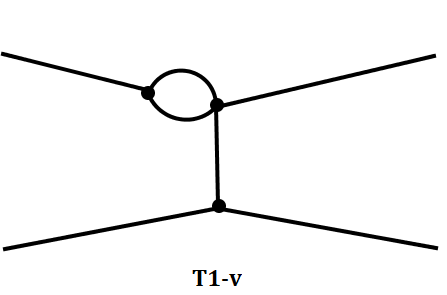}\hspace{0.12in}
\includegraphics[scale=0.35]{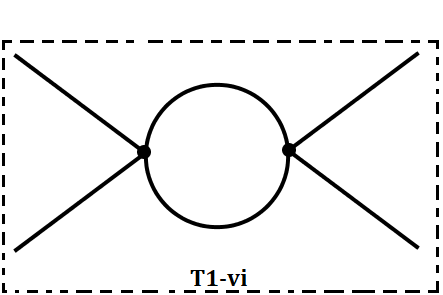}
$$
\caption{ All possible topologies at the one-loop with four external legs. The topology inside the black dashed box leads to non-renormalizable models.  }\label{one-a}
\end{figure}

Without further ado, we now conduct a systematic search to find minimal one-loop models. 
Following the diagrammatic approach, one can find out all possible one-loop topologies associated with four external legs \cite{Bonnet:2012kz}, which are listed in Fig. \ref{one-a} after discarding the self-energy-like diagrams.
However, not all the topologies listed in   Fig. \ref{one-a} can lead to successful one-loop neutrino masses due to various reasons  to be discussed below.  First, note that  in our scenario, two of the external legs must be fermions for any of the topologies listed above. As a result, 
the topology T1-vi,  corresponds to non-renormalizable operator, hence we eliminate it immediately.  

\FloatBarrier
\begin{figure}[th!]
\centering
$$
\includegraphics[scale=0.4]{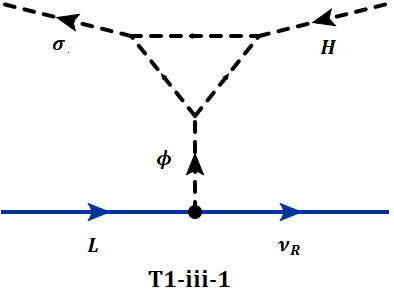}\hspace{0.1in}
\includegraphics[scale=0.4]{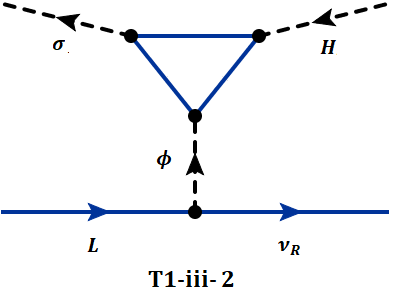}\hspace{0.1in}
\includegraphics[scale=0.4]{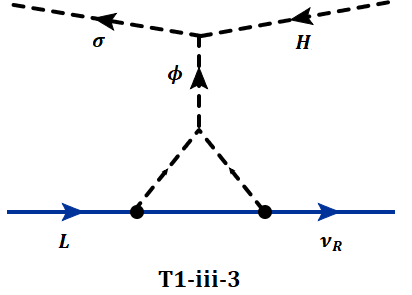}
$$
$$
\includegraphics[scale=0.4]{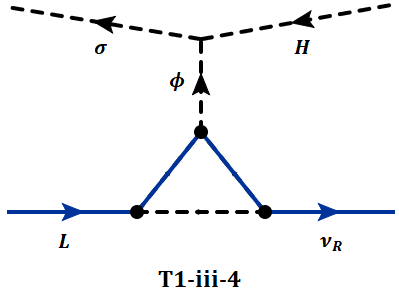}\hspace{0.1in}
\includegraphics[scale=0.4]{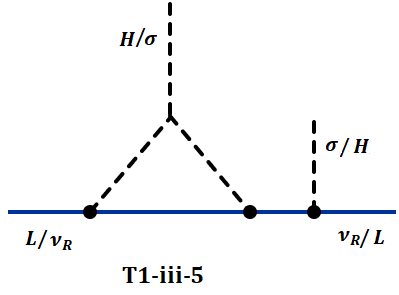}\hspace{0.1in}
\includegraphics[scale=0.4]{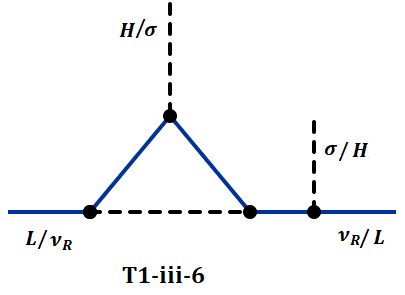}
$$
\caption{ One loop diagrams arising from topology T1-iii. }\label{one-b}
\end{figure}

Till now the Lorentz nature of each line is not specified. To proceed further we specify the spinor and scalar nature of the propagators.  First, we focus  on the T1-iii topology,  which leads to six different diagrams that can be categorized as: (a) diagram with all three internal fermion (or scalar) propagators, (b)  diagram with  two internal fermion  propagators, (c)  diagram where only one of the three internal states is a fermion. All these six possible diagrams are presented in Fig. \ref{one-b}. From these diagrams,  one immediately sees that T1-iii-y with y=1-4, all contain the  same scalar $\phi$ attached to the external fermion (scalar) lines in diagrams T1-iii-1,2 (T1-iii-3,4).   In all these models, regardless of the other particles circulating in the loop,  the quantum number of this scalar is uniquely fixed to be  $\phi (1,2,-\frac{1}{2},3)$ by either the external fermion or scalar fields. As a result, in the Lagrangian,  a cubic term of the form $\mathcal{L}\supset H \phi \epsilon \sigma^{\ast} \supset H^0\phi^0\sigma^{0\ast}$ is allowed, which  leads to an induced VEV of the neutral component of $\phi$  after both the $U(1)_{B-L}$ and EW symmetries are broken. Consequently, neutrinos receive a tree-level mass from the $d=4$ renormalizable term: $\mathcal{L}\supset y\; \overline{L}\phi\nu_R$ that is invariant under the entire gauge symmetry, SM$\times U(1)_{B-L}$. Hence we discard these four\footnote{It is worthwhile to point out that depending on the model,  fermionic mass insertions or scalar mixings are required inside the loop for these diagrams to be convergent.} diagrams T1-iii-y with y=1-4. 

\FloatBarrier
\begin{figure}[th!]
\centering\includegraphics[height=0.3\textheight,width=0.9\textwidth]{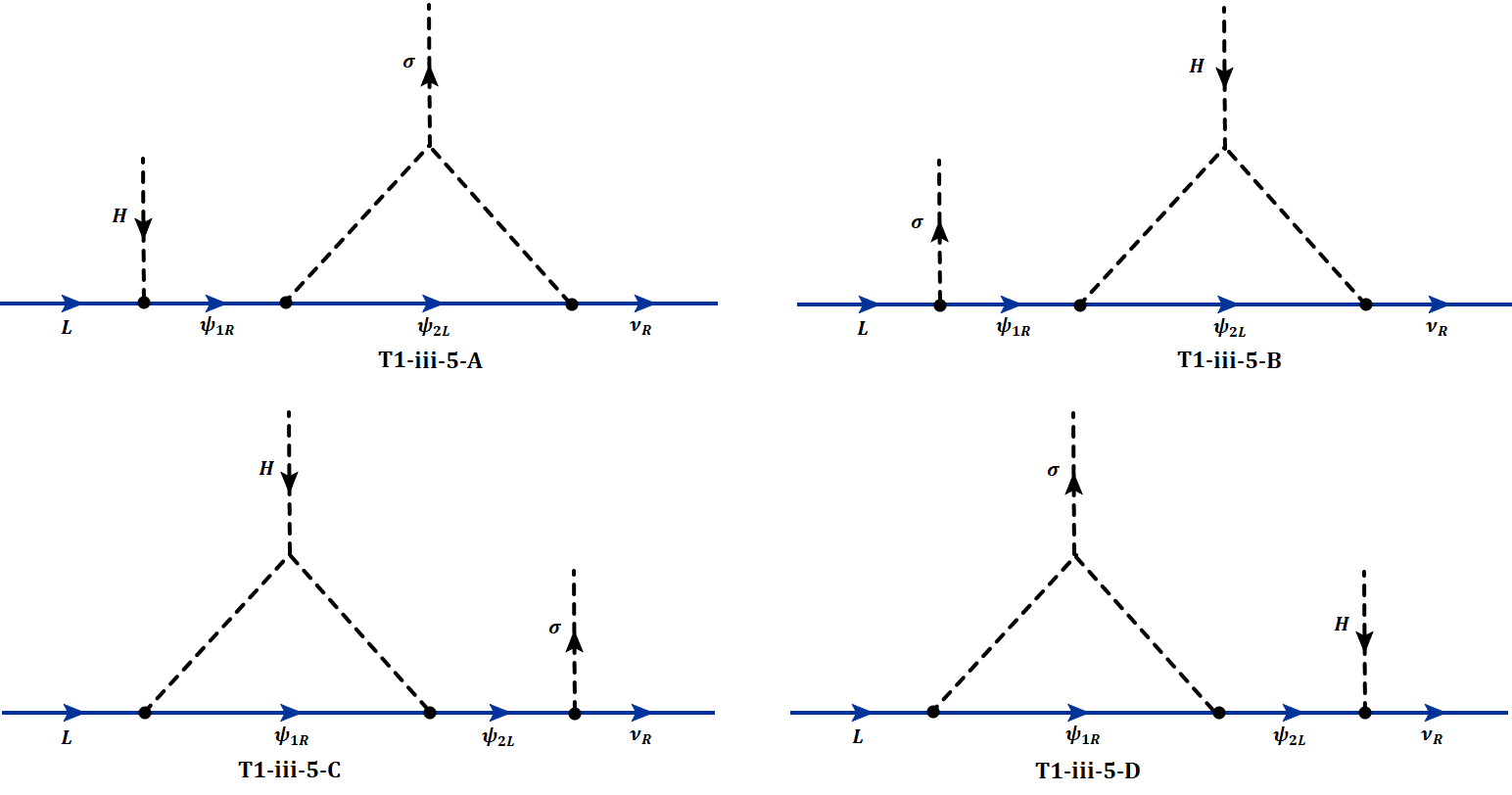}\hspace{0.2in}
\caption{ Variations of the one loop diagrams of the type T1-iii-5.  }\label{one-c}
\end{figure}

Before discussing the T1-iii-5,6 possibilities, here we leave a brief remark cornering  topologies T1-iv,v. For these topologies to provide renormalizable models, the lower two external particles must be fermions.  In such a scenario,   it is now straightforward from the above discussion to realize that constructing models out of these two topologies demand the same scalar field $\phi (1,2,-1/2,3)$, hence generate neutrino mass at tree-level and fail to qualify to be a true radiative model.

\begin{figure}[th!]
\centering\includegraphics[height=0.1\textheight,width=0.9\textwidth]{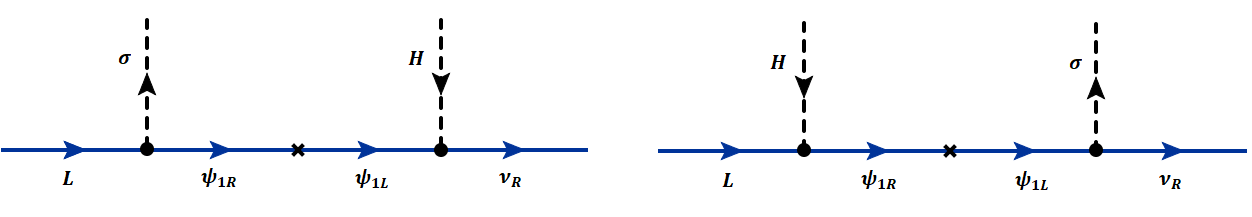}\hspace{0.2in}
\caption{ Tree-level Dirac seesaw reduction from diagrams of type T1-iii-3.  }\label{dirac-seesaw}
\end{figure}

Let us now discuss the second last  diagram  T1-iii-5 listed in Fig.  \ref{one-b}. From this, different variations of diagrams  can be constructed by swapping the external Higgs doublet and singlet which  are presented in Fig. \ref{one-c}. It is straightforward to try to build models associated with  each of these four types of diagrams. However, here we show that none of these diagrams leads  to any successful radiative  model. To get an understanding, first, consider the T1-iii-5-A model diagram. From the leftmost vertex, the quantum number of the fermion fields ${\psi_1}_{L,R}$ is automatically fixed to be $(1,1,0,-1)$, without specifying the quantum numbers of the rest of the particles in the loop.  Introduction of such a fermion leads to the following terms in the Lagrangian: 
\begin{align}
\mathcal{L}\supset y_1\overline{L}\widetilde{H}{\psi_1}_{R} + y_2 \overline{{\psi_1}}_{L} \sigma \nu_R + M_{1}\overline{{\psi_1}}_{L}{\psi_1}_R.     
\end{align}
The existence of these terms is responsible to give tree-level mass to the neutrinos via Dirac seesaw as shown in Fig. \ref{dirac-seesaw} (left diagram). 
It is not difficult to see that the model diagram T1-iii-5-B demands the existence of a fermion ${\psi_1}\sim (1,2,-\frac{1}{2},-4)$, which allows the following terms in the Lagrangian: 
\begin{align}
\mathcal{L}\supset y_1^{\prime}\overline{L}\sigma{\psi_1}_{R} + y_2^{\prime} \overline{{\psi_1}}_{L} \widetilde{H} \nu_R + M_{1}^{\prime}\overline{{\psi_1}}_{L}{\psi_1}_R.     
\end{align}
Hence, also leads to tree-level Dirac seesaw as demonstrated in Fig. \ref{dirac-seesaw} (right diagram). Following similar arguments, diagrams  T1-iii-5-C and T1-iii-5-D must also be discarded.  The same conclusion can be reached for T1-iii-6 diagram as shown in Fig. \ref{one-b}.  

\FloatBarrier

%%%%%%%%%%%%%%%%%%%%%%%%%%%%%%%%%%%%%%%%%%%%%%%
\subsection{Search for minimal one-loop models without  colored particles}
\begin{figure}[th!]
\centering
$$
\includegraphics[scale=0.4]{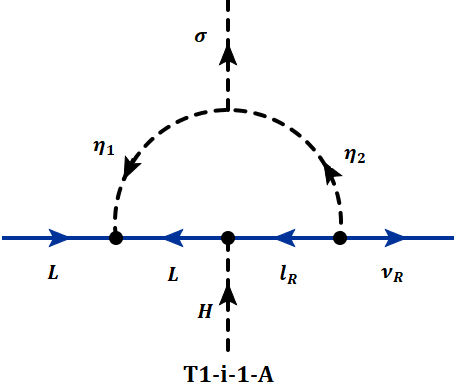}\hspace{0.12in}
\includegraphics[scale=0.4]{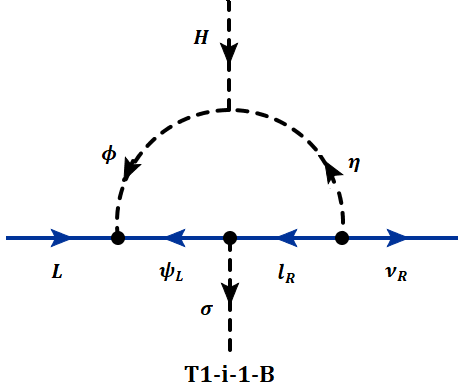}\hspace{0.12in}
\includegraphics[scale=0.4]{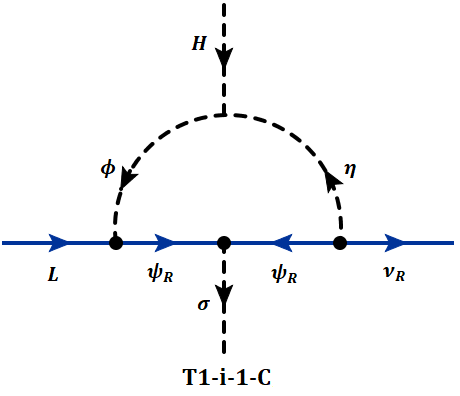}
$$
$$
\includegraphics[scale=0.4]{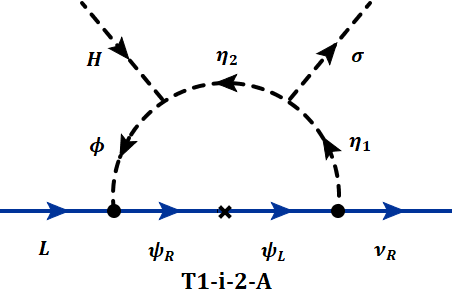}\hspace{0.12in}
\includegraphics[scale=0.4]{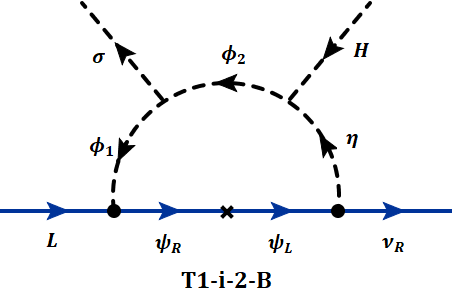}\hspace{0.12in}
\includegraphics[scale=0.4]{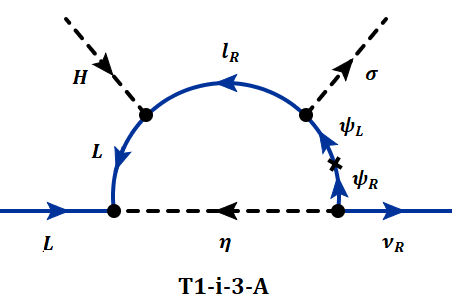}
$$
$$
\includegraphics[scale=0.4]{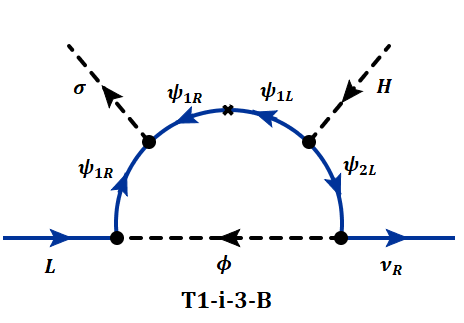}\hspace{0.12in}
\includegraphics[scale=0.4]{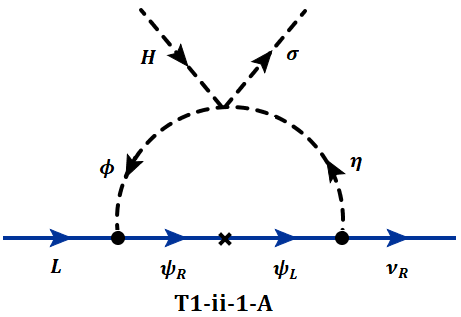}
$$
\caption{ One-loop model diagrams emerging from T1-i and T1-ii topologies in our framework.   Quantum numbers associated with each of the models are presented in Table \ref{table-one}. Notation: the color-blind scalars $\eta$ and $\phi$ are assumed to be singlet and doublet under $SU(2)_L$.}\label{one-d}
\end{figure}

From the aforementioned elaborated discussion, we exclude topologies T1-iii,iv,v,vi and in the following build successful minimal models arising from the remaining two topologies, T1-i and T1-ii.  Affiliated with topology T1-i, there are three different diagrams corresponding to three different Lorentz structures that we denote by T1-i-1,2,3. Within each of these diagrams,  interchanging the two external scalar legs gives two distinct possibilities. Altogether these six possible  diagrams are presented in Fig. \ref{one-d} along with the unique diagram arising from T1-ii topology.   Below, we fabricate economical models from these seven diagrams. Quantum numbers of all the particles associated with each of these models are summarized in Table \ref{table-one}.    

\FloatBarrier

\begin{itemize}
    \item[$\square$] \underline{T1-i-1-A}:
\end{itemize}

It is pointed out in Ref. \cite{Saad:2019bqf} that the most economical model at the  one-loop for Dirac neutrino mass in the framework we are working is
associated with the T1-i-1-A model diagram.   This minimal model consists of only two additional singly charged scalars that are iso-singlets and carry non-zero $B-L$ charges: $\eta_1(1,1,1,2)$ and $\eta_2(1,1,1,5)$. One of the main reasons  T1-i-1-A is the  simplest among the rest of the one-loop models is, neither any exotic fermion beyond the SM nor any scalar charged under $SU(2)_L$ needs to be introduced. 

\begin{itemize}
    \item[$\square$] \underline{T1-i-1-B/C}:
\end{itemize}

The difference between the T1-i-1-A and T1-i-1-B/C model diagrams is just the swapping of the $H$ and $\sigma$ scalars. Since, $\sigma$ is now attached to the internal fermion line instead of the scalar line, more possibilities open up. The most economical models that can be built out of these, require a pair of vector-like Dirac fermions  $\psi_{L,R}\sim (1,1,Y,\alpha)$ in addition to two scalars. In T1-i-1-B model diagram, since one of the internal fermions is $\ell_R$, one requires $Y=-1$ and $\alpha=-4$. As a result, one of the BSM scalars must be a singly charged state and the other an iso-doublet with $3/2$ hypercharge. However, it should be kept in mind that such a  model (T1-i-1-B) is somewhat phenomenologically constrained. The reason is, it requires the fermionic state $\psi_{L,R}\sim (1,1,-1,-4)$ and contains the following terms in {the} Lagrangian:
\begin{align}
\mathcal{L}\supset \begin{pmatrix} \overline{L}&\overline{\psi_L} \end{pmatrix}    
\begin{pmatrix} y_eH&y\sigma\\y^{\prime}\sigma^{\ast}&M_{\psi} \end{pmatrix}  
\begin{pmatrix} \ell_R\\\psi_R \end{pmatrix}.  \label{mix}
\end{align}
When $U(1)_{B-L}$ and EW symmetries are broken, these terms lead to $6\times 6$ mass matrix for the SM leptons and the vector-like leptons. The mixing among these fields must be small  to satisfy all the collider constraints provided that $\psi$ has mass $\sim \mathcal{O}$(TeV), which requires the off-diagonal Yukawa couplings $y, y^{\prime}$ to be  small in Eq. \eqref{mix}.

On the other hand, if only the BSM fermion is inside the  loop,  then the vertex $\psi_R\psi_R\sigma^{\ast}$  associated with the  model diagram T1-i-1-C demands $Y=0$ and $\alpha=3/2$. Consequently, one of the BSM Higgs is SM singlet and the second one is SM like Higgs doublet with non-zero $B-L$ charge. Quantum numbers corresponding to T1-i-1-B/C model diagrams are written down in Table  \ref{table-one} and the associated Feynman diagrams are presented in Fig. \ref{one-d}.

\begin{itemize}
    \item[$\square$] \underline{T1-i-2-A/B}:
\end{itemize}

The minimal models that are built from T1-i-2-A  and T1-i-2-B  both require extension by one pair of fermions $\psi_{L,R} (1,1,Y,\alpha)$ along with three scalars. The corresponding models presented in Table \ref{table-one} show that among the three scalars, the former model contains two singlets and one iso-doublet, whereas, the latter model consists of  two  iso-doublets and a singlet, all carrying non-zero $B-L$ charge. Note that for T1-i-2-A and T1-i-2-B model diagrams,  if $Y=0$ then $\alpha\neq -1$ must be realized, otherwise they lead to tree-level Dirac seesaw to generate neutrino mass. Also to avoid direct tree-level mass term for the case of  T1-i-2-A (T1-i-2-B), if $Y=-1$ or $Y=0$ then   $\alpha\neq -7, -5, 2, 5$ ($\alpha\neq -10, -7, -1, 2$)  must be satisfied to forbid the induced VEV of the multiplet $\phi$.

\begin{table}[!h]
\centering
%\label{oneloop}
\begin{center}
 \resizebox{1\textwidth}{!}{\begin{tabular}{|| c | c | c | c | c | c || }
\hline

\textbf{Diagram} & \textbf{Models}  &\multicolumn{2}{ c |}{\textbf{New Fields}}  & \textbf{Relevant terms in Lagrangian} & \textbf{New ?}   \\ 
\cline{3-4}
 & & \textbf{Scalars} & \textbf{Fermions} &  &   \\
\hline
T1-i-1 & \textbf{T1-i-1-A} & \parbox[c]{4.7cm}{\vspace{2pt}$\eta_1(1,1,1,2)$\\ $\eta_2(1,1,1,5)$\vspace{2pt}} & --- & \parbox[c]{4.7cm}{\vspace{2pt}$y_1 \overline{L^c}\epsilon \eta_1 L 
+ y_e \overline{L} H l_R $\\ $+ y_2 \overline{l^c_R} \eta_2 \nu_R + \mu \eta_2 \eta_1^* \sigma^*$\vspace{2pt}}  &  \cite{Calle:2018ovc,Saad:2019bqf}  \\
  \cline{2-6}
 & \textbf{\textbf{T1-i-1-B}} & \parbox[c]{4.7cm}{\vspace{2pt}$\eta(1,1,1,5)$\\$\phi(1,2,\frac{3}{2},5)$\vspace{2pt}} & \parbox[c]{4.7cm}{\vspace{2pt}$\psi_{L,R} (1,1,-1,-4)$\vspace{2pt}} & \parbox[c]{4.7cm}{\vspace{2pt}$y_1 \overline{L^c}\epsilon \phi \psi_L + y_2 \overline{\psi_L} \sigma^* l_R $\\ $ + y_3 \overline{l^c_R} \eta \nu_R + \mu \phi^{*} H  \eta$\vspace{2pt}} &  \color{MyGreen}{ $\cmark$} \\
 \cline{2-6}
  & \textbf{\textbf{T1-i-1-C}} & \parbox[c]{4.7cm}{\vspace{2pt}$\eta(1,1,0,\frac{5}{2})$\\ $\phi(1,2,\frac{1}{2},\frac{5}{2})$\vspace{2pt}} & \parbox[c]{4.7cm}{\vspace{2pt}$\psi_{L,R} (1,1,0,\frac{3}{2})$\vspace{2pt}}  & \parbox[c]{4.7cm}{\vspace{2pt}$y_1 \overline{L}\epsilon \phi^* \psi_R + y_2 \overline{\psi^c_R} \sigma^* \psi_R $\\ $ + y_3 \overline{\psi^c_R} \eta \nu_R + \mu \phi^{*} H \eta$\vspace{2pt}} &  \color{MyGreen}{ $\cmark$}  \\
\hline
T1-i-2 &\textbf{T1-i-2-A} & \parbox[c]{4.7cm}{\vspace{2pt}$\eta_1 (1,1,Y,4+\alpha)$\\ $\eta_2 (1,1,Y,1+\alpha)$\\ $\phi (1,2,Y+\frac{1}{2},1+\alpha)$\vspace{2pt}} & \parbox[c]{4.7cm}{\vspace{2pt}$\psi_{L,R} (1,1,Y,\alpha)$\vspace{2pt}} & \parbox[c]{4.7cm}{\vspace{2pt}$ M_{\psi}\overline{\psi_L}\psi_R+y_1 \overline{L}\epsilon \phi^* \psi_R $\\$+ y_2 \overline{\psi_L} \eta_1 \nu_R + \mu_1 \eta_1 \eta_2^*\sigma^* $\\$ + \mu_2 \phi^{*} H \eta_2$\vspace{2pt}} &  \color{MyGreen}{ $\cmark$} \\
 \cline{2-6}
 & \textbf{T1-i-2-B}  & \parbox[c]{4.7cm}{\vspace{2pt}$\eta(1,1,Y,4+\alpha)$\\$\phi_1 (1,2,Y+\frac{1}{2},1+\alpha)$\\$\phi_2 (1,2,Y+\frac{1}{2},4+\alpha)$\vspace{2pt}} & \parbox[c]{4.7cm}{\vspace{2pt}$\psi_{L,R}(1,1,Y,\alpha)$\vspace{2pt}} &\parbox[c]{4.7cm}{\vspace{2pt}$ M_{\psi}\overline{\psi_L}\psi_R+y_1 \overline{L}\epsilon \phi_1^* \psi_R $\\$+ y_2 \overline{\psi_L} \eta \nu_R + \mu_1 H \eta \phi_2^* $\\$ + \mu_2 \phi^{*}_1 \phi_2 \sigma^*$\vspace{2pt}} &  \color{MyGreen}{ $\cmark$} \\
\hline
T1-i-3 & \textbf{T1-i-3-A} & \parbox[c]{4.7cm}{\vspace{2pt}$\eta(1,1,1,2)$\vspace{2pt}}  & \parbox[c]{4.7cm}{\vspace{2pt}$\psi_{L,R} (1,1,-1,2)$\vspace{2pt}}  & \parbox[c]{4.7cm}{\vspace{2pt}$ M_{\psi}\overline{\psi_L}\psi_R+y_1 \overline{L^c}\epsilon \eta L  $\\$+ y_2 \overline{\psi_R^c} \eta \nu_R + y_3 \overline{\psi_L}\sigma l_R $\\$ + y_e \overline{L}H l_R$\vspace{2pt}}  &  \color{MyGreen}{ $\cmark$}  \\
 \cline{2-6} 
 & \textbf{T1-i-3-B} &  \parbox[c]{4.7cm}{\vspace{2pt}$\phi (1,2,\frac{1}{2},\frac{5}{2})$\vspace{2pt}} &  \parbox[c]{4.7cm}{\vspace{2pt}$\psi_{1L,R} (1,1,0,\frac{3}{2})$\\$\psi_{2L,R} (1,2,\frac{1}{2},-\frac{3}{2})$\vspace{2pt}} & \parbox[c]{4.7cm}{\vspace{2pt}$ M_{\psi_1}\overline{\psi_{1L}}\psi_{1R} + y_1 \overline{L}\epsilon \phi^* \psi_{1R}$\\$ + y_2 \overline{\psi^c_{1R}} \sigma^* \psi_{1R} + y_3 \overline{\psi_{2L}} \phi \nu_R $\\$ + y_4 \overline{\psi^c_{2L}} H^* \psi_{1L} $\vspace{2pt}}  &   \color{MyGreen}{ $\cmark$} \\
\hline 
T1-ii-1 & \textbf{T1-ii-1-A} & \parbox[c]{4.7cm}{\vspace{2pt}$\eta (1,1,Y,4+\alpha)$\\$ \phi (1,2,Y+\frac{1}{2},1+\alpha)$\vspace{2pt}} & \parbox[c]{4.7cm}{\vspace{2pt}$\psi_{L,R} (1,1,Y,\alpha)$\vspace{2pt}} & \parbox[c]{4.7cm}{\vspace{2pt}$ M_{\psi}\overline{\psi_L}\psi_R + y_1 \overline{L}\epsilon \phi^* \psi_R $\\$ 
+ y_2 \overline{\psi_L} \eta \nu_R + \lambda \phi^{*} H \eta \sigma^*$\vspace{2pt}} &  \cite{Calle:2018ovc,Bonilla:2018ynb}  \\
\hline  
\end{tabular}}
\end{center}\caption{ Minimal one-loop models without colored particles constructed from $d=5$ effective operator given in Eq. \eqref{d5}. Notation: the color-blind scalars $\eta$ and $\phi$ are assumed to be singlet and doublet under $SU(2)_L$.  A term of the form $\phi^{\ast}H$ in the  Lagrangian refers to $\phi^{\dagger}H$, and furthermore, $\epsilon H^{\ast}\equiv \widetilde{H}$ and $\epsilon \phi^{\ast}\equiv \widetilde{\phi}$.  }\label{table-one}
\end{table}

\begin{itemize}
    \item[$\square$] \underline{T1-i-3-A/B}:
\end{itemize}

To build viable models out of T1-i-3-A one needs a pair of BSM fermions $\psi\sim (1,1,-1,2)$ and a singly charged scalar $(1,1,1,2)$. Whereas, this model is very economical, but the viability of the   T1-i-3-B model diagram, obtained by flipping $H$ with $\sigma$   requires two pairs of fermions: an iso-singlet $\psi_{1}(1,1,0,3/2)$  and an iso-doublet $\psi_2(1,2,1/2,-3/2)$ along with a scalar doublet $\phi(1,2,1/2,5/2)$. 

\begin{itemize}
    \item[$\square$] \underline{T1-ii-1-A}:
\end{itemize}

The last in our list for the one-loop models is the one derived from T1-ii topology, which is unique in structure since, interchanging the two external Higgs lines do not lead to different possibilities.  The associated model diagram  T1-ii-1-A  is presented in Fig. \ref{one-d}. The minimal model constructed from this diagram requires an iso-doublet scalar, an iso-singlet scalar and a pair of iso-singlet fermions $\psi\sim (1,1,Y,\alpha)$ listed in Table \ref{table-one}. Here, to avoid any direct mass term between the left-handed and the right-handed neutrinos, for $Y=-1$, $\alpha\neq 2$ must be realized. Furthermore, to forbid tree-level Dirac seesaw, if $Y=0$, $\alpha\neq -1$ constraint must be imposed as well.   

At this point, it is important to note that any model that is built from diagram T1-i-2 will always lead to a second Feynman diagram which is exactly the same as T1-ii-1, even though these two sets of diagrams  originate from completely different topologies. From Fig. \ref{one-d} one sees that in diagram T1-i-2, removing the scalar  that is propagating in between the two scalar cubic vertices   immediately gives a second diagram which is the same as T1-i-2.  In this sense, unlike the rest of the models presented here, models emerging from  T1-i-2-A,B diagrams cannot be considered as genuine and, we refer these two as \textit{non-genuine} model diagrams. Here \textit{non-genuinity} does not refer to one-loop  to tree-level reduction.

%%%%%%%%%%%%%%%%%%%%%%%%%%%%%%%%%%%%%%%%%%%%%%%
\subsection{Search for minimal one-loop models with colored particles}
\begin{figure}[th!]
\centering
$$
\includegraphics[scale=0.4]{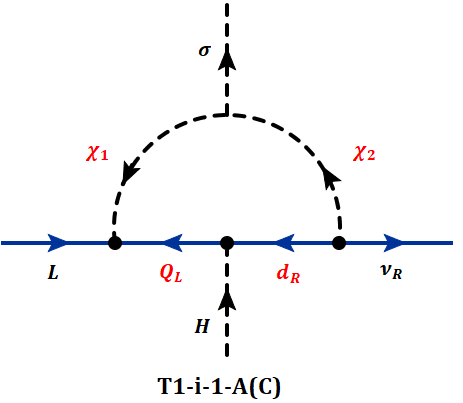}\hspace{0.1in}
\includegraphics[scale=0.4]{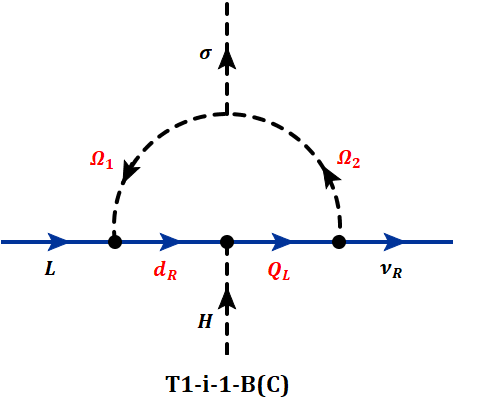}\hspace{0.1in}
\includegraphics[scale=0.4]{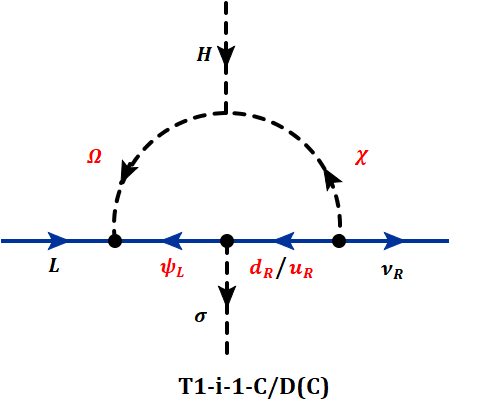}
$$
$$
\includegraphics[scale=0.4]{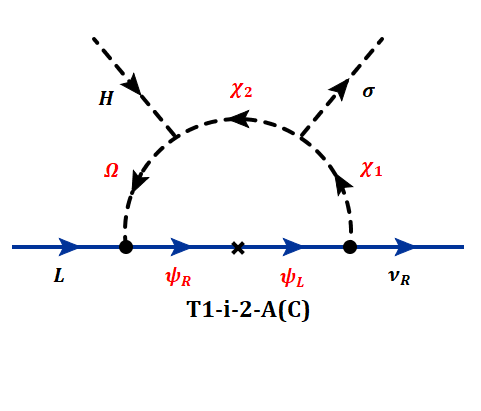}\hspace{0.1in}
\includegraphics[scale=0.4]{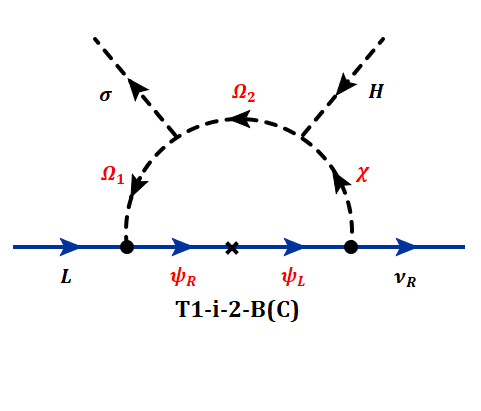}
$$
$$
\includegraphics[scale=0.4]{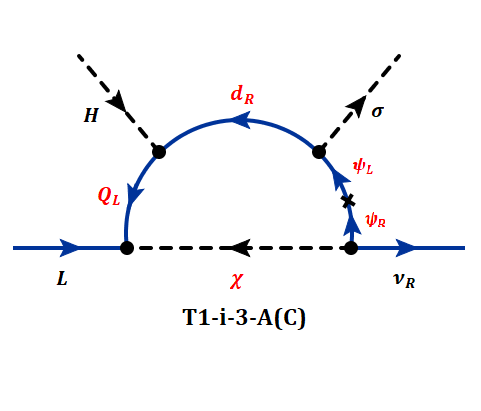}\hspace{0.1in}
\includegraphics[scale=0.4]{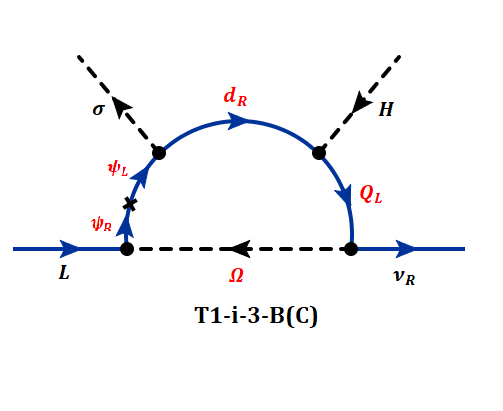}\hspace{0.1in}
\includegraphics[scale=0.4]{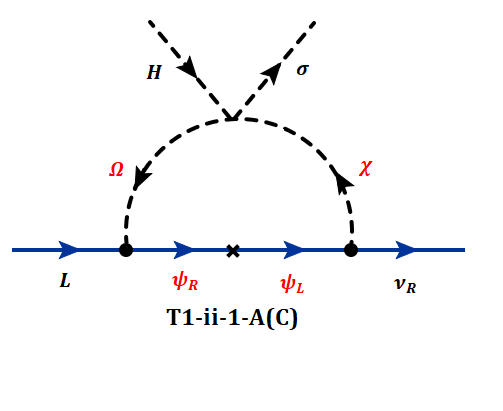}
$$
\caption{Color version of one loop diagrams emerging from T1-i and T1-ii topologies. Quantum numbers associated with each of the model diagrams are presented in Table \ref{table-one}. Notation: the  scalars,  $\chi$ and $\Omega$ charged under the color group are assumed to be $SU(2)_L$ singlet and doublet, respectively. }\label{one-e}
\end{figure}

 As discussed in Sec. \ref{SEC-01}, we also include colored particles in our search. In this section, we
build the minimal models with BSM  states running in the loop that are charged under $SU(3)_C$. Following the above discussions,  here we  only consider the  topologies that can provide viable models. Whereas it is straightforward to build the 
 associated colored versions of the models, however, for some of the cases, models with and without colored particles differ in structure.  Here we very briefly discuss the models and point out the differences when required. Successful minimal model diagrams with colored particles are presented in  Fig. \ref{one-e} and their quantum numbers are listed in Table \ref{table-one-color}. To distinguish these models, we label the model diagrams by T1-x-y-z(C),   here C in the parentheses is put to differentiate models with colored particles, compared to models with color singlets. 
 
\begin{itemize}
    \item[$\square$] \underline{T1-i-1-A/B/C/D(C)}:
\end{itemize}
 
The model diagram T1-i-1-A(C) is straightforwardly  obtained from 
T1-i-1-A by replacing the two singly charged scalars $\eta_1 (1,1,1,2)$ $+ \eta_2 (1,1,1,5)$ by two scalar  leptoquarks $\chi_1 (\overline{3},1,1/3,2/3)$  $+ \chi_2 (\overline{3},1,1/3,11/3)$. This choice is uniquely fixed by making the replacements: $L\leftrightarrow Q_L$ and $\ell_R \leftrightarrow d_R$ inside loop in T1-i-1-A to obtain T1-i-1-A(C). Just like T1-i-1-A, this model again can be considered as the most minimal radiative Dirac neutrino mass model with colored particles.  Note that a straightforward variation of T1-i-1-A(C), is the interchange $Q_L\leftrightarrow d_R$, which gives rise to the model diagram T1-i-1-B(C). It is worthwhile to mention that no such analogous model with color singlets exists. It is due to fact that $L\leftrightarrow \ell_R$ exchange would give rise to a model which would automatically permit tree-level mass for neutrinos.   Even though the model associated to T1-i-1-B(C) also contains only two scalar leptoquarks, but  is less economical compared to T1-i-1-A(C) in the sense that both these  leptoquarks are in the fundamental representation of  $SU(2)_L$. However, both these models are very economical since, no BSM fermionic extension is required.  

Two different colored version models can be constructed that are analogous to T1-i-1-B by replacing $\ell_R \leftrightarrow d_R$ (T1-i-1-C(C)) and $\ell_R \leftrightarrow u_R$ (T1-i-1-D(C)) respectively. As noted above, the  T1-i-1-B model is  phenomenologically somewhat constraining, the same  argument follows for T1-i-1-C(C) and T1-i-1-D(C) due to the mixing of the SM fermions with BSM fermionic states.  
Note however that analogous to T1-i-1-C model diagram with color singlet states, no such colored version model can be constructed with a single pair of BSM fermionic states. 

\begin{itemize}
    \item[$\square$] \underline{T1-i-2-A/B(C), T1-i-3-A/B(C) $\&$ T1-ii-1-A(C)}:
\end{itemize}

As already pointed out, models fabricated from diagram T1-i-2 are non-genuine and always contain a second diagram which is exactly the same as T1-ii-1 that emerge from a completely different topology and more economical in nature.  Allowing color  non-singlet states do not alter this fact. Colored version of these models can be straightforwardly built by following the corresponding discussions of the associated models with color singlets.

Whereas the colored extension of T1-i-3-A(C) from  T1-i-3-A is straightforward, construction of T1-i-3-B(C) deserves further explanation. As aforementioned, the model built out of  T1-i-3-B  requires the introduction of two pairs of BSM fermions. It is interesting to note that, the associated colored version, T1-i-3-B(C) is comparatively economical as implementation of this minimal model diagram can be done with just one pair BSM fermions $\psi\sim (3,1,-1/3,10/3)$.  Besides this fermion,  just one more BSM Higgs is employed to complete the loop. For more details on these models  see Fig. \ref{one-e} and Table \ref{table-one-color}.

\FloatBarrier

\begin{table}[!h]
\centering
%\label{oneloop}
\begin{center}
 \resizebox{1\textwidth}{!}{\begin{tabular}{|| c | c | c | c | c | c || }
\hline

\textbf{Diagram} & \textbf{Models}  &\multicolumn{2}{ c |}{\textbf{New Fields}}  & \textbf{Relevant terms in Lagrangian} & \textbf{New ?}   \\ 
\cline{3-4}
 & & \textbf{Scalars} & \textbf{Fermions} &  &   \\
\hline
T1-i-1 & \textbf{T1-i-1-A(C)} & \parbox[c]{4.7cm}{\vspace{2pt}$\textcolor{red}{\chi_1}(\bar{3},1,\frac{1}{3},\frac{2}{3})$\\ $\textcolor{red}{\chi_2}(\bar{3},1,\frac{1}{3},\frac{11}{3})$\vspace{2pt}} & --- & \parbox[c]{4.7cm}{\vspace{2pt}$y_1 \overline{L^c}\epsilon \chi_1 Q_L 
+ y_d \overline{Q_L} H d_R $\\ $+ y_2 \overline{d^c_R} \chi_2 \nu_R + \mu \chi_2 \chi_1^* \sigma^*$\vspace{2pt}} &  \color{MyGreen}{ $\cmark$} \\
 \cline{2-6} 
  & \textbf{T1-i-1-B(C)}  &  \parbox[c]{4.7cm}{\vspace{2pt}$\textcolor{red}{\Omega_1}(3,2,\frac{1}{6},\frac{4}{3})$\\$\textcolor{red}{\Omega_2}(3,2,\frac{1}{6},\frac{13}{3})$\vspace{2pt}} & --- & \parbox[c]{4.7cm}{\vspace{2pt}$y_1 \overline{d_R}\Omega_1 \epsilon L 
+ y_d \overline{Q_L} H d_R $\\ $+ y_2 \overline{Q_L} \Omega_2 \nu_R + \mu \Omega_1^{*} \Omega_2 \sigma^*$\vspace{2pt}} &  \color{MyGreen}{ $\cmark$} \\
 \cline{2-6}
  & \textbf{\textbf{T1-i-1-C(C)}} & \parbox[c]{4.7cm}{\vspace{2pt}$\textcolor{red}{\chi} (\bar{3},1,\frac{1}{3},\frac{11}{3})$\\$\textcolor{red}{\Omega}(\bar{3},2,\frac{5}{6},\frac{11}{3})$\vspace{2pt}} & \parbox[c]{4.7cm}{\vspace{2pt}$\textcolor{red}{\psi_{L,R}} (3,1,-\frac{1}{3},-\frac{8}{3})$ \vspace{2pt}}&\parbox[c]{4.7cm}{\vspace{2pt}$y_1 \overline{L^c}\epsilon \Omega \psi_L
+ y_2 \overline{\psi_L} \sigma^* d_R $\\ $+ y_3 \overline{d^c_R} \chi \nu_R + \mu \Omega^{*} H \chi $\vspace{2pt}} &  \color{MyGreen}{ $\cmark$} \\
\cline{2-6}  
 & \textbf{\textbf{T1-i-1-D(C)}} &  \parbox[c]{4.7cm}{\vspace{2pt}$\textcolor{red}{\chi} (\bar{3},1,-\frac{2}{3},\frac{11}{3})$ \\ $\textcolor{red}{\Omega}(\bar{3},2,-\frac{1}{6},\frac{11}{3})$\vspace{2pt}}  & \parbox[c]{4.7cm}{\vspace{2pt}$\textcolor{red}{\psi_{L,R}} (3,1,\frac{2}{3},-\frac{8}{3})$ \vspace{2pt}}& \parbox[c]{4.7cm}{\vspace{2pt}$y_1 \overline{L^c}\epsilon \Omega \psi_L
+ y_2 \overline{\psi_L} \sigma^* u_R $\\ $+ y_3 \overline{u^c_R} \chi \nu_R + \mu \Omega^{*} H \chi $\vspace{2pt}} &  \color{MyGreen}{ $\cmark$} \\
\hline
T1-i-2 &  \textbf{T1-i-2-A(C)} & \parbox[c]{4.7cm}{\vspace{2pt}$\textcolor{red}{\chi_1} (3,1,Y,4+\alpha)$\\$\textcolor{red}{\chi_2} (3,1,Y,1+\alpha)$ \\$\textcolor{red}{\Omega} (3,2,Y+\frac{1}{2},1+\alpha)$\vspace{2pt}} & \parbox[c]{4.7cm}{\vspace{2pt} $\textcolor{red}{\psi_{L,R}} (3,1,Y,\alpha)$\vspace{2pt}}& \parbox[c]{4.7cm}{\vspace{2pt}$ M_{\psi} \overline{\psi_L}\psi_R + y_1 \overline{L}\epsilon \Omega^* \psi_R $\\ $+ y_2 \overline{\psi_L} \chi_1 \nu_R + \mu_1 \Omega^{*} H \chi_2 $\\$+ \mu_2 \chi_1 \chi_2^* \sigma^*$\vspace{2pt}} &  \color{MyGreen}{ $\cmark$} \\
 \cline{2-6}
  &  \textbf{T1-i-2-B(C)} & \parbox[c]{4.7cm}{\vspace{2pt}$\textcolor{red}{\chi}(3,1,Y,4+\alpha)$\\ $\textcolor{red}{\Omega_1} (3,2,Y+\frac{1}{2},1+\alpha)$\\ $\textcolor{red}{\Omega_2} (3,2,Y+\frac{1}{2},4+\alpha)$\vspace{2pt}} & \parbox[c]{4.7cm}{\vspace{2pt}$\textcolor{red}{\psi_{L,R}} (3,1,Y,\alpha)$\vspace{2pt}} & \parbox[c]{4.7cm}{\vspace{2pt}$ M_{\psi} \overline{\psi_L}\psi_R + y_1 \overline{L}\epsilon \Omega_1^* \psi_R $\\ $+ y_2 \overline{\psi_L} \chi \nu_R + \mu_1 \Omega_2^{*} H \chi $\\$+ \mu_2 \Omega_1^{*} \Omega_2 \sigma^*$\vspace{2pt}} &  \color{MyGreen}{ $\cmark$} \\
\hline
T1-i-3 & \textbf{T1-i-3-A(C)} & \parbox[c]{4.7cm}{\vspace{2pt}$\textcolor{red}{\chi} (\bar{3},1,\frac{1}{3},\frac{2}{3})$\vspace{2pt}} & \parbox[c]{4.7cm}{\vspace{2pt}$\textcolor{red}{\psi_{L,R}} (3,1,-\frac{1}{3},\frac{10}{3})$\vspace{2pt}} & \parbox[c]{4.7cm}{\vspace{2pt}$ M_{\psi} \overline{\psi_L}\psi_R + y_1 \overline{L^c}\epsilon \chi Q_L 
$\\$+ y_d \overline{Q_L} H d_R + y_2 \overline{\psi^c_R} \chi \nu_R $\\ $ + y_3 \overline{\psi_L} \sigma d_R$\vspace{2pt}} &  \color{MyGreen}{ $\cmark$} \\ 
 \cline{2-6}
  &  \textbf{T1-i-3-B(C)} &\parbox[c]{4.7cm}{\vspace{2pt}$\textcolor{red}{\Omega} (3,2,\frac{1}{6},\frac{13}{3})$\vspace{2pt}} &\parbox[c]{4.7cm}{\vspace{2pt}$\textcolor{red}{\psi_{L,R}} (3,1,-\frac{1}{3},\frac{10}{3})$\vspace{2pt}} & \parbox[c]{4.7cm}{\vspace{2pt}$ M_{\psi} \overline{\psi_L}\psi_R + y_1 \overline{L}\epsilon \Omega^* \psi_R $\\$+ y_d \overline{Q_L} H d_R + y_2 \overline{Q_L} \Omega \nu_R $\\ $ + y_3 \overline{\psi_L} \sigma d_R$\vspace{2pt}} &  \color{MyGreen}{ $\cmark$} \\
\hline 
T1-ii-1 & \textbf{T1-ii-1-A(C)} & \parbox[c]{4.7cm}{\vspace{2pt}$\textcolor{red}{\chi} (3,1,Y,4+\alpha)$ \\ $\textcolor{red}{\Omega} (3,2,Y+\frac{1}{2},1+\alpha)$ \vspace{2pt}} & \parbox[c]{4.7cm}{\vspace{2pt}$\textcolor{red}{\psi_{L,R}}(3,1,Y,\alpha)$\vspace{2pt}} & \parbox[c]{4.7cm}{\vspace{2pt}$ M_{\psi} \overline{\psi_L}\psi_R + y_1 \overline{L}\epsilon \Omega^* \psi_R 
$\\$+ y_2 \overline{\psi_L} \chi \nu_R + \lambda \Omega^{*} H \sigma^* \chi$\vspace{2pt}} &  \color{MyGreen}{ $\cmark$} \\
\hline  
\end{tabular}}
\end{center}\caption{ Minimal one-loop models with colored particles constructed from $d=5$ effective operator given in Eq. \eqref{d5}. Notation: the  scalars,  $\chi$ and $\Omega$ charged under the color group are assumed to be $SU(2)_L$ singlet and doublet, respectively.}\label{table-one-color}
\end{table}
\clearpage

%%%%%%%%%%%%%%%%%%%%%%%%%%%%%%%%%%%%%%%%%%%%%%%
%%%%%%%%%%%%%%%%%%%%%%%%%%%%%%%%%%%%%%%%%%%%%%%
\section{Search for minimal  two-loop models}\label{SEC-04}
In the previous section, we have systematically built the minimal one-loop Dirac neutrino mass models within our framework. In this section, our goal is to implement a similar methodology to select  the minimal two-loop models. 
%%%%%%%%%%%%%%%%%%%%%%%%%%%%%%%%%%%%%%%%%%%%%%%
\subsection{Generic topologies, nomenclature and  selecting viable skeleton diagrams}
\begin{figure}[th!]
\centering\includegraphics[height=0.4\textheight,width=0.8\textwidth]{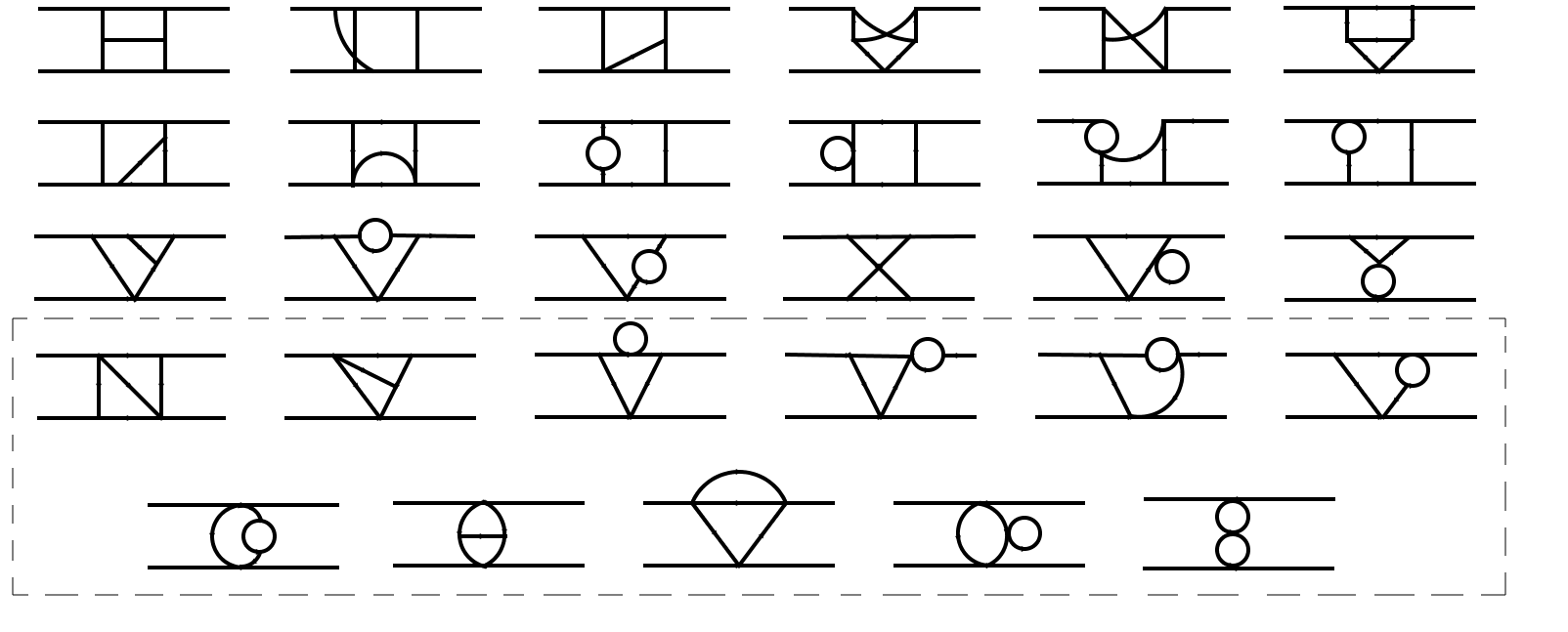}\hspace{0.2in}
\caption{ All two-loop topologies arising from $d=5$ effective operator Eq. \eqref{d5}.  }\label{two-a}
\end{figure}

Following a similar diagrammatic  approach as before, all possible two-loop topologies with four external legs can be  identified. Excluding the tadpole diagrams,  one-particle-reducible two-loop topologies and by removing the  topologies involving self-energies in the external legs,  in total $29$ distinct possibilities are identified in Ref. \cite{Sierra:2014rxa} and for completeness we present them in Fig. \ref{two-a}. In this list, the last $11$ of them correspond to non-renormalizable topologies, hence must be  discarded immediately.   

\FloatBarrier

\begin{figure}[th!]
\centering\includegraphics[scale=0.33]{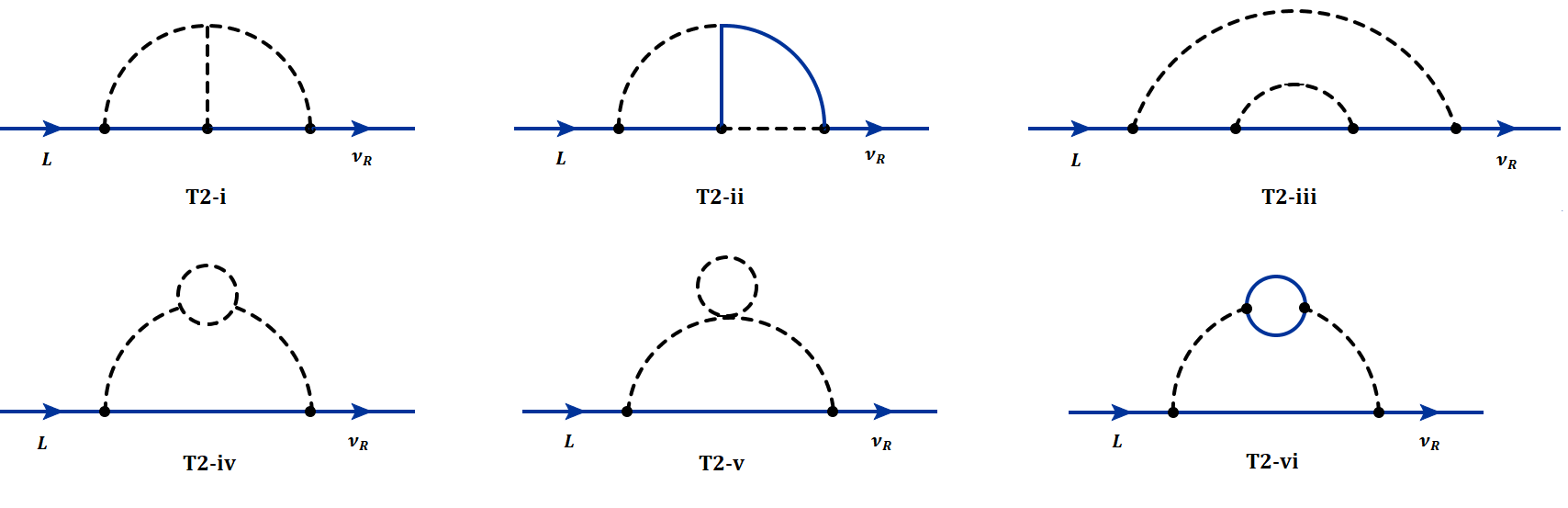}\hspace{0.2in}
\caption{ Generic two-loop skeleton diagrams arising from $d=5$ effective operator Eq. \eqref{d5} from which we build the minimal models. See text for  details.  }\label{two-b}
\end{figure}

While dealing with  the remaining 18 two-loop topologies, the number of possible diagrams emerging as a result of specifying the spinors and scalars is quite large compared to the one-loop scenario.  Furthermore, when all different swapping of the external legs are also taken into account, a vast number of diagrams appear within our framework because, unlike the Majorana scenario, two of the external fermions are not identical anymore. The same thing is also true for the two external scalar legs, they are not identical.  The appearance of the immense  number of diagrams is  also partly because of the singlet nature of one of the external scalars, which has more freedom to be attached with the internal propagators.   In the following, we will demonstrate this by providing explicit examples.
Interestingly, all these topologies can be reduced to only six basic diagrams once the external scalar legs are suppressed \cite{Sierra:2014rxa, CentellesChulia:2019xky}. 
To make our analysis more tractable, 
instead of starting from each of these 18 possible   renormalizable topologies of Fig. \ref{two-a}, we  focus on the set of these six basic diagrams derived from the generic two-loop topologies.  These are the general diagrams resulting from removing the external scalar legs, and we call them the \textit{skeleton diagrams}, which within our working framework have the structures as  shown in Fig. \ref{two-b}.   However, each of these skeleton diagrams may lead to quite a few numbers of viable diagrams depending on how the external scalar legs are attached, which makes the analysis of the two-loop models somewhat tedious compared to one-loop case.
So the way we proceed is, we start with each of the  skeleton diagrams given in  Fig. \ref{two-b}, and systematically  identify  only the most economical model diagram by following our minimality principles as aforementioned.  Below we discuss how this filtering process is done at length. 

\FloatBarrier

The convention we follow is: each of the  skeleton diagrams given in Fig. \ref{two-b} is labelled with T2-x. Here 2 represents two-loop for obvious reason and x= i - vi to differentiate among the  skeleton diagrams from each other. As mentioned above, each of these skeleton diagrams can lead to multiple diagrams after inserting the external scalar legs  that we denote by T2-x-y  (with y=1, 2, 3, ...). Furthermore, models  that are built from a diagram T2-x-y will be named as    T2-x-y-z (where z=A, B, C, ...), that is  T2-x-y-z represents a model diagram where the quantum number of all the particles are specified. If the internal particles carry color charge, the model diagram will be named  T2-x-y-z(C), here C in the parentheses represents color.

%%%%%%%%%%%%%%%%%%%%%%%%%%%%%%%%%%%%%%%%%%%%%%%
\subsection{Search for minimal two-loop models without  colored particles}
We now proceed to build  the associated minimal models from each of the skeleton diagrams. 

\begin{itemize}
    \item[$\square$] \underline{T2-i-y-A}:
\end{itemize}

\begin{figure}[th!]
\centering\includegraphics[scale=0.36]{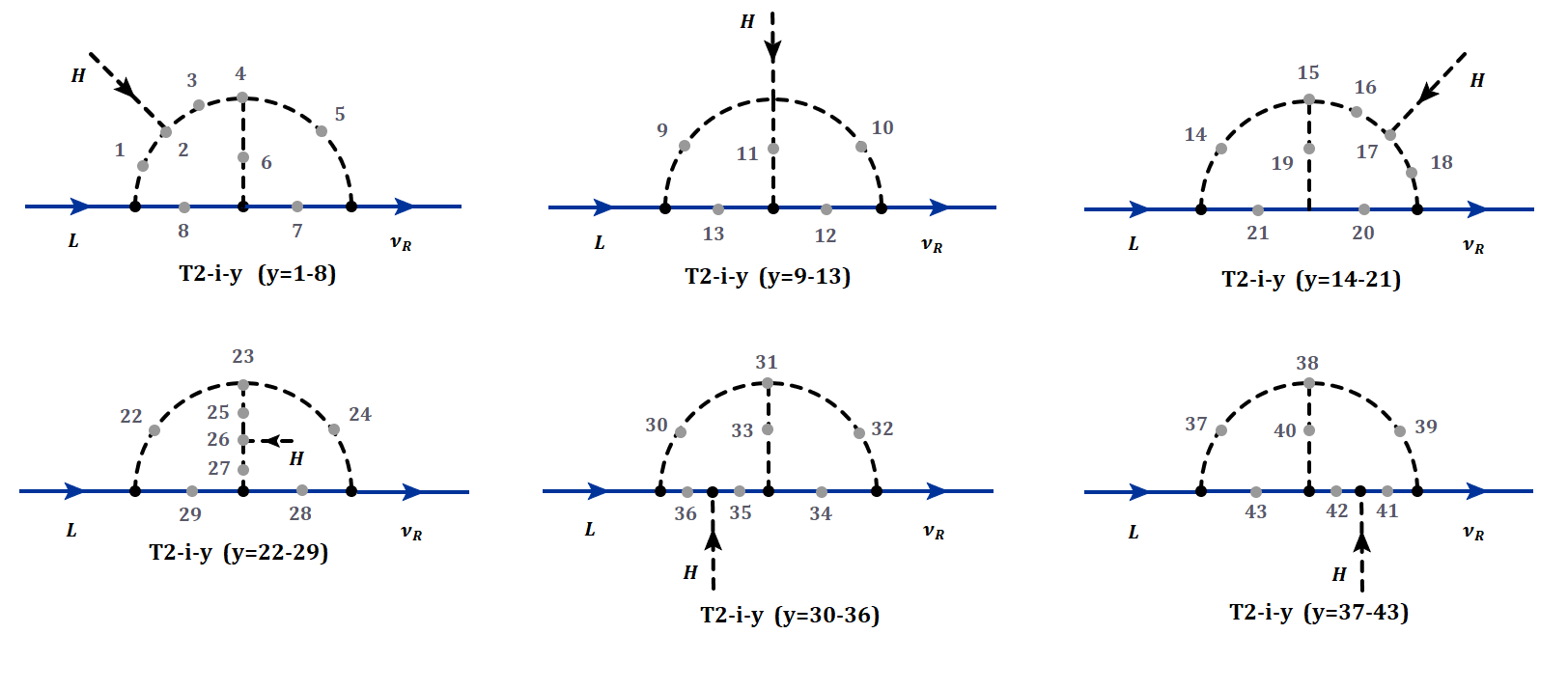}\hspace{0.2in}
\caption{ All possible two-loop diagrams manufactured  from T2-i skeleton diagram. The gray colored dots  represent the different ways of linking $\sigma$ to the internal propagators, leading to different diagrams T2-i-y, where y= 1 - 43.  }\label{two-c}
\end{figure}

First, we consider the T2-i skeleton diagram from Fig. \ref{two-b} and systemically construct all possible diagrams emerging from it and provide our arguments in singling out the most economical model diagram. Note that there are six different ways the SM Higgs doublet  $H$ can be connected with the internal propagators.  Furthermore, corresponding to each of these six cases, the scalar singlet, $\sigma$ can be attached in multiple different ways. By repeating this process for all of these six scenarios, we end up with a total of 43 viable diagrams, T2-i-y (y=1-43) as explicitly demonstrated in Fig. \ref{two-c}.  In each of these figures, the gray-colored dots represent the different ways of linking $\sigma$ to the internal propagators. Each of these different possibilities corresponds to a distinct model.  

\FloatBarrier

\begin{figure}[th!]
\centering
$$
\includegraphics[scale=0.4]{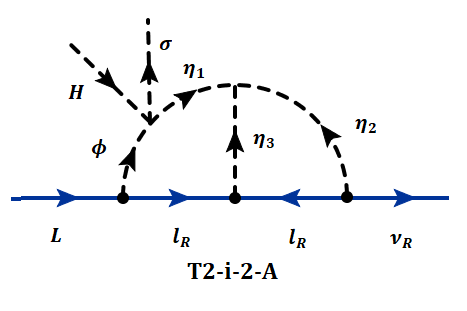}\hspace{0.03in}
\includegraphics[scale=0.4]{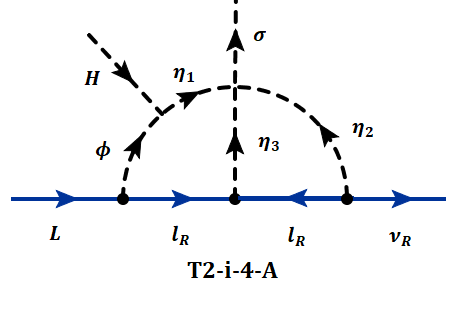}\hspace{0.03in}
\includegraphics[scale=0.4]{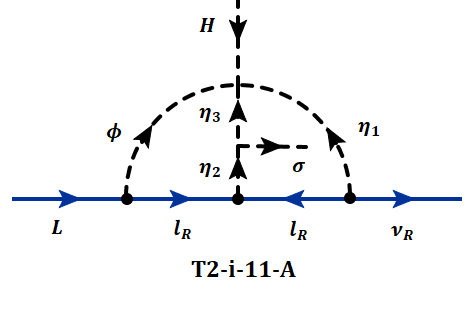}
$$
$$
\includegraphics[scale=0.4]{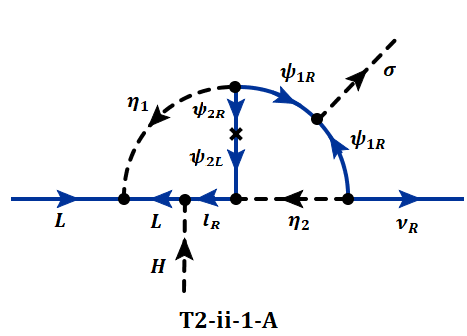}\hspace{0.1in}
\includegraphics[scale=0.4]{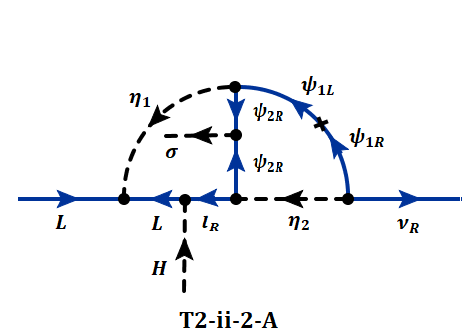}\hspace{0.1in}
\includegraphics[scale=0.4]{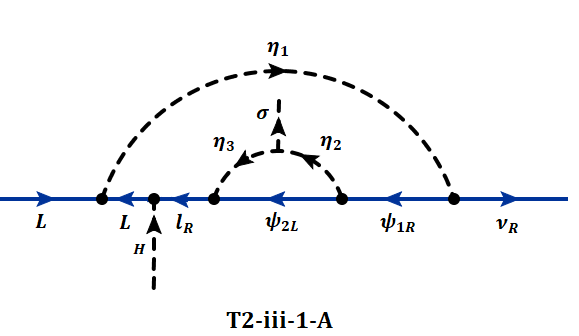}\hspace{0.1in}
$$
$$
\includegraphics[scale=0.4]{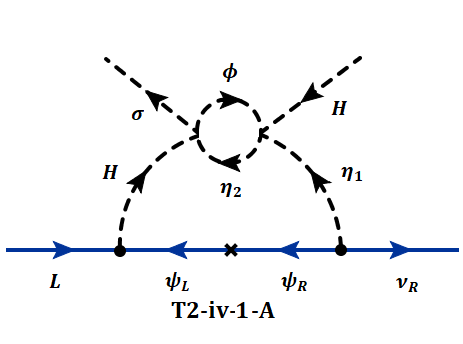}\hspace{0.1in}
\includegraphics[scale=0.4]{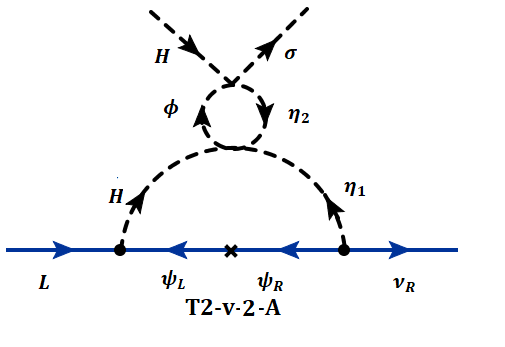}\hspace{-0.06in}
\includegraphics[scale=0.41]{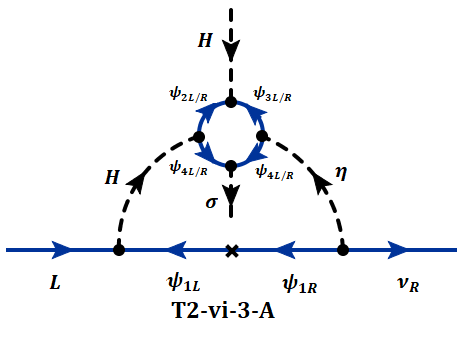}
$$
\caption{ Minimal  two-loop model diagrams arising from T2-x (x=i-vi) skeleton diagrams. Quantum numbers associated with each of the model diagrams are presented in Table \ref{table-two}. Notation: the color-blind scalars $\eta$ and $\phi$ are assumed to be singlet and doublet under $SU(2)_L$. }\label{two-e}
\end{figure}

As aforementioned, we are not going to search for  models corresponding to each of these diagrams, rather our focus is to  pick  the most minimal model diagram originating from T2-i. Going toward this direction,  we immediately discard 22 viable diagrams T2-i-y with y=
7-8, 12-13, 20-21, 28-29, 34-36, 41-43. The reason is that to build a model from each of these diagrams, requires the introduction of  BSM fermions and does not fall into the minimal category.  For the rest of the diagrams, in search for finding the minimal model, we fix the internal fermions to be $\ell_R$, consequently, 7 of the diagrams T2-i-y with y= 2-6 and 10-11,  require a second Higgs doublet (with non-zero $B-L$ charge), rest of the diagrams demand more number of scalar iso-doublets for the completion of the loops,  hence do not satisfy our minimality  postulates.  For these remaining  7 diagrams, we select the models with a minimum number of additional scalars on top of the second iso-doublet, and once this counting is done,  we construct the associated explicit model diagrams. Only three of the model diagrams pass our minimality axioms, which are T2-i-2,4,11 and are  shown in   Fig. \ref{two-e}.
 Besides the second iso-doublet,  the  models associated with  T2-i-2,4 (T2-i-11) composed of  two singly (doubly) charged and a doubly (singly) charged scalars.  For the complete quantum numbers of these particles, see Table \ref{table-two}.

From the exercise done above, one can understand that finding the minimal model starting from a certain skeleton diagram can be a bit tedious process. Following the same procedure  for the rest of the skeleton diagrams presented in Fig. \ref{two-b}, we summarize the most economical models in Table \ref{table-two}. In the following, we briefly discuss this systematic analysis in search  of the  minimal models for the rest of the skeleton diagrams.   

\begin{itemize}
    \item[$\square$] \underline{T2-ii-1/2-A}:
\end{itemize}

Now consider the skeleton diagram T2-ii. Just like the previous example, it is not difficult to see that there are many different ways for the $H$ and $\sigma$ fields to be attached with the particles running in the loop. Instead of presenting all these possible ways of realizing two-loop neutrino mass models arising from the T2-ii skeleton diagram, we only limit ourselves to the  scenario which is the most economical.  For the sake of minimality, we choose to insert the SM Higgs doublet in the internal fermion line which is closest to the external $L$ leg.  As a result, only BSM iso-singlet representations can be employed to complete the loop diagram, hence satisfies our minimality constraints.  Furthermore, to reduce the number of additional scalars,  we attach the singlet Higgs $\sigma$  with one of the internal fermion lines, which however, does not increase the number of required BSM fermion multiplets. 
The first model (labelled by T2-ii-1-A) we manufacture here is by attaching $\sigma$  to the fermion propagator that is closest to $\nu_R$ external leg. 
With these choices, one needs two pairs of additional iso-singlet fermions and two BSM iso-singlet scalars to complete the loop diagram. A slight variation of this model can be made (labelled by T2-ii-2-A) by inserting $\sigma$ in the internal fermion line which is not connected to either of the external fermion legs.  The associated model diagrams are presented in Fig. \ref{two-e} and the quantum numbers of all the needed particles are listed in Table \ref{table-two}. 
 
\begin{itemize}
    \item[$\square$] \underline{T2-iii-1-A}:
\end{itemize}

To build an economical model from skeleton diagram T2-iii, where different variations of the scalar field  insertions are again available,   
one can follow the arguments that are not very different from the previous models constructed. To aim to find the minimal model, in this particular scenario, we attach the SM Higgs doublet with the internal fermion line\footnote{If instead, $H$ is attached to the outer scalar propagator, completion of the diagram would require one BSM scalar doublet, which can be considered as the next-to-minimal model of similar category, however belong to a different model-diagram. } that is closest to the incoming $L$ field. As a result, no BSM scalar doublet needs to be employed in this construction. Furthermore, we connect the external  singlet scalar with the inner scalar loop as shown in  Fig. \ref{two-e}. This choice of inserting the external scalar fields guarantees the requirement of minimum number of scalar and fermion states, as well as  demands only iso-singlet states.   
Looking at the details of the quantum numbers of the particle contents of this model listed in Table \ref{table-two}, one finds that for $Y=0$, $\alpha\neq -1$ and $\alpha\neq -7, 4$ must be imposed to avoid tree-level Dirac seesaw and one-loop reduction respectively.   

\FloatBarrier

\begin{itemize}
    \item[$\square$] \underline{T2-iv-1-A}:
\end{itemize}

Here we discuss the skeleton diagram T2-iv. It will be shown that the skeleton diagrams T2-x, with x=iv-vi are very special in nature and share some common features.  Building models out of these three cases are very much constraining due to the special nature.  First, we discuss the construction of T2-iv in more detail, the analogous arguments can be used to build models from T2-v,vi. 

\begin{figure}[th!]
\centering\includegraphics[scale=0.44]{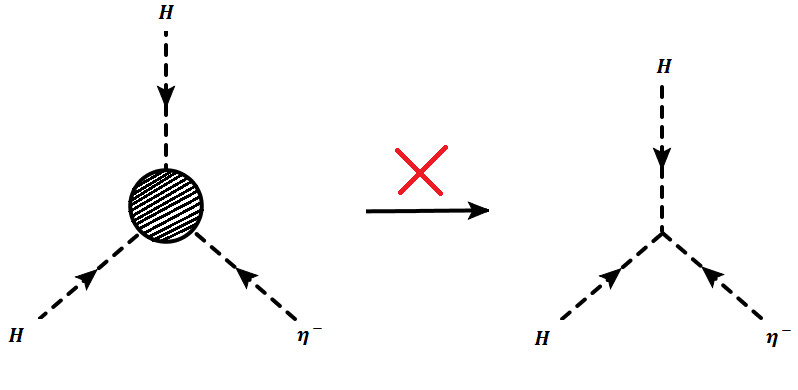}\hspace{0.2in}
\caption{ Required type of the internal scalar vertex correction for skeleton diagrams of type T2-iv, T2-v and  T2-vi. }\label{two-g}
\end{figure}

\begin{figure}[th!]
\centering\includegraphics[scale=0.43]{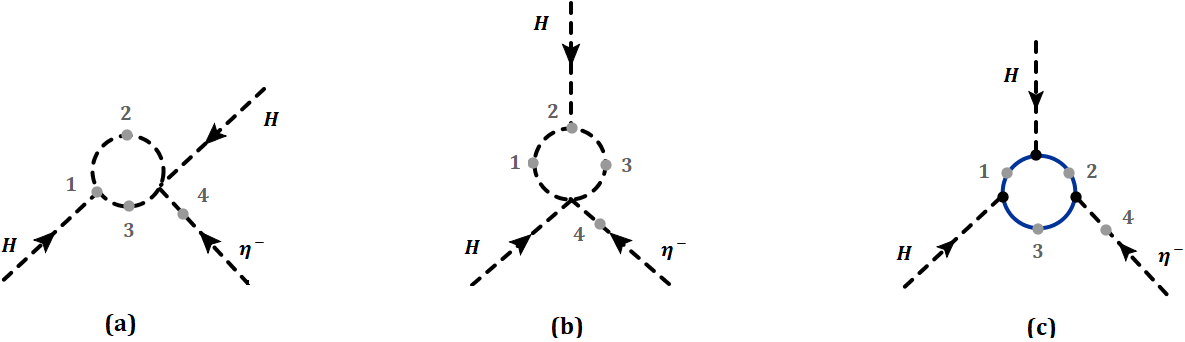}\hspace{0.2in}
\caption{ Possible ways of inserting $\sigma$ field   for skeleton diagrams of type T2-iv, T2-v and  T2-vi. See text for details.   }\label{two-h}
\end{figure}

The only viable type of model diagrams that can be formed out of T2-iv must contain the effective scalar vertex of the form $H H \eta$ (with $\eta\sim (1,1,-1)$ under the SM) as shown in see Fig. \ref{two-g}, here the blob must contain a loop to qualify  to be a true two-loop model. The reason for this requirement is, if instead two different Higgs doublets ($H$ and $\phi$) are allowed to form this vertex, then the loop contained in the blob immediately reduces to a tree-vertex of the form $H \phi \eta$, hence becomes a one-loop model. On the other hand, if the same SM Higgs is playing the role, then the tree-level vertex automatically vanishes  as a result of the   antisymmetric property of the term $\mathcal{L}\supset \mu H \epsilon H \eta^-$. Hence, loop to tree-level vertex reduction is not allowed as demonstrated in Fig. \ref{two-g}. With this restriction, still, there are
freedoms to attach the external $H$ with the internal particles contained by the blob.  To reduce the number of required additional iso-doublets, we make use of a quartic coupling that consists of external $H$ and internal $\eta$ with two more BSM scalars as shown in Fig \ref{two-h}-a. Moreover, we have not used the freedom of inserting $\sigma$ yet, which in fact can be done in four different ways as depicted in Fig.   \ref{two-h}-a, T2-iv-y, with y=1-4. Out of these four choices, y=1 certainly requires the minimum number of additional scalars which is just two. Consequently, we only consider this economical choice and discard the rest of the possibilities.  Between these two scalars, one of them is an iso-doublet with non-zero $B-L$ charge $\phi\sim (1,2,Y,\alpha)$. Note that as aforementioned, if $Y=-1/2$, $\alpha\neq 3$ must be realized to forbid direct mass term for neutrinos at the tree-level.  In addition to the scalars, these models require a pair of vector-like Dirac fermions having similar quantum numbers as the $e^c$ under the SM, hence these fermions need to have a mass of order $\mathcal{O}$(TeV) or higher to be consistent with phenomenological observation as mentioned above.   

\FloatBarrier

\FloatBarrier

\begin{itemize}
    \item[$\square$] \underline{T2-v-1-A}:
\end{itemize}

The above arguments can be repeated for fabricating models from the T2-v  skeleton diagram. Even though for this scenario, there is no freedom in inserting external $H$ leg, however, a limited number of freedom is still available associated with $\sigma$, which are  demonstrated in Fig.   \ref{two-h}-b.  Out of these four scenarios, we pick T2-v-2 to build the most economical model which is presented in Fig.  \ref{two-e} and the quantum numbers of all the needed particles are listed in Table \ref{table-two}.

\begin{itemize}
    \item[$\square$] \underline{T2-vi-3-A}:
\end{itemize} 

The last in our list is the T2-vi skeleton diagram, which requires quite a few BSM fermions to complete the loop diagram. Unlike T2-iv, no such quartic scalar vertex is allowed here, hence there is no freedom in connecting $H$. Though, there are four different ways one  can attach $\sigma$ as depicted in Fig. \ref{two-h}-c, minimality  corresponds to T2-vi-3 realization. This choice requires four BM fermions and a singly charged scalar, for details see Table \ref{table-two}.

Even though T2-x (x=iv-vi)   diagrams  appear to provide divergent loop integrals, but it is not the case. The reason is, when the term $H\epsilon H$ is broken down to its component fields, two same terms emerge with opposite sign, removing the divergences from the difference diagrams resulting in finite neutrino mass, for details see Ref.  \cite{Sierra:2014rxa}. Consequently, models arising from T2-iv,v,vi are very special in nature.

\begin{table}[!h]
%\label{twoloop}
\begin{center}
\resizebox{1.0\textwidth}{!}{\begin{tabular}{|| c | c | c | c | c | c || }
\hline
\textbf{Skeleton} & \textbf{Models}  &\multicolumn{2}{ c |}{\textbf{New Fields}}  & \textbf{Relevant terms in Lagrangian} & \textbf{New ?}   \\ 
\cline{3-4}
 & & \textbf{Scalars} & \textbf{Fermions} &  &   \\
\hline
T2-i & \textbf{T2-i-2-A} & \parbox[c]{4.7cm}{\vspace{2pt}$\eta_1 (1,1,1,-3)$\\ $\eta_2 (1,1,1,5)$\\ $\eta_3 (1,1,-2,-2)$\\ $\phi (1,2,\frac{1}{2},0)$\vspace{2pt}} & --- & \parbox[c]{4.7cm}{\vspace{2pt}$y_1 \overline{L} \phi l_R 
+ y_2 \overline{l^c_R} \eta_3^* l_R $\\ $+ y_3 \overline{l^c_R} \eta_2 \nu_R + \mu \eta_1 \eta_2 \eta_3 $\\$+ \lambda \phi \epsilon H \sigma^* \eta_1^*  $\vspace{2pt}}  &  \cite{Saad:2019bqf} \\
  \cline{2-6}
 & \textbf{T2-i-4-A} & \parbox[c]{4.7cm}{\vspace{2pt}$\eta_1 (1,1,1,0)$\\ $\eta_2 (1,1,1,5)$\\ $\eta_3 (1,1,-2,-2)$\\$\phi (1,2,\frac{1}{2},0)$\vspace{2pt}} & --- & \parbox[c]{4.7cm}{\vspace{2pt}$y_1 \overline{L} \phi l_R 
+ y_2 \overline{l^c_R} \eta_3^* l_R $\\ $+ y_3 \overline{l^c_R} \eta_2 \nu_R + \mu \phi \epsilon H \eta_1^* $\\$+ \lambda \eta_1 \eta_2 \eta_3 \sigma^*   $\vspace{2pt}} &  \color{MyGreen}{ $\cmark$}  \\
 \cline{2-6}
 & \textbf{T2-i-11-A} & \parbox[c]{4.7cm}{\vspace{2pt}$\eta_1 (1,1,1,5)$\\$\eta_2 (1,1,-2,-2)$\\ $\eta_3 (1,1,-2,-5)$\\ $\phi (1,2,\frac{1}{2},0)$\vspace{2pt}} &--- & \parbox[c]{4.7cm}{\vspace{2pt}$y_1 \overline{L} \phi l_R 
+ y_2 \overline{l^c_R} \eta_2^* l_R $\\ $+ y_3 \overline{l^c_R} \eta_1 \nu_R + \mu \eta_2 \eta_3^* \sigma^* $\\$+ \lambda \phi \epsilon H \eta_1 \eta_3 $\vspace{2pt}} &  \color{MyGreen}{ $\cmark$}  \\
\hline
T2-ii & \textbf{T2-ii-1-A} & \parbox[c]{4.7cm}{\vspace{2pt}$\eta_1 (1,1,1,2)$\\ $\eta_2 (1,1,0,\frac{5}{2})$\vspace{2pt}} &   \parbox[c]{4.7cm}{\vspace{2pt}$\psi_{1L,R} (1,1,0,\frac{3}{2})$\\$\psi_{2L,R} (1,1,-1,-\frac{7}{2})$\vspace{2pt}} & \parbox[c]{4.7cm}{\vspace{2pt}$  M_{\psi_2}\overline{\psi_{2L}}\psi_{2R} + y_1 \overline{L^c} \epsilon \eta_1 L $\\$ 
+ y_2 \overline{\psi_{2L}} \eta_2^* l_R + y_3 \overline{\psi^c_{1R}} \eta_2 \nu_R $\\ $ + y_4 \overline{\psi^c_{1R}} \eta_1 \psi_{2R} + y_5 \overline{\psi^c_{1R}} \sigma^* \psi_{1R} $\\$+ y_e \overline{L} H l_R $\vspace{2pt}} & \color{MyGreen}{ $\cmark$}  \\
\cline{2-6}
& \textbf{T2-ii-2-A} & \parbox[c]{4.7cm}{\vspace{2pt}$\eta_1 (1,1,1,2)$\\ $\eta_2 (1,1,-1,\frac{1}{2})$\vspace{2pt}} &   \parbox[c]{4.7cm}{\vspace{2pt}$\psi_{1L,R} (1,1,1,\frac{7}{2})$\\$\psi_{2L,R} (1,1,0,\frac{3}{2})$\vspace{2pt}} & \parbox[c]{4.7cm}{\vspace{2pt}$  M_{\psi_1}\overline{\psi_{1L}}\psi_{1R} + y_1 \overline{L^c} \epsilon \eta_1 L $\\$ + y_2 \overline{\psi^c_{2R}} \eta_2^* l_R + y_3 \overline{\psi^c_{1R}} \eta_2 \nu_R $\\ $ + y_4 \overline{\psi_{1L}} \eta_1 \psi_{2R} + y_5 \overline{\psi^c_{2R}} \sigma^* \psi_{2R} $\\$+ y_e \overline{L} H l_R $ \vspace{2pt}} & \color{MyGreen}{ $\cmark$}  \\
\hline
T2-iii & \textbf{T2-iii-1-A} & \parbox[c]{4.7cm}{\vspace{2pt}$\eta_1 (1,1,-1,-2)$\\ $\eta_2 (1,1,-(Y+1),2-\alpha)$\\ $\eta_3 (1,1,-(Y+1),-(1+\alpha))$\vspace{2pt}} & \parbox[c]{4.7cm}{\vspace{2pt}$\psi_{1L,R} (1,1,-1,2)$\\$\psi_{2L,R} (1,1,Y,\alpha)$\vspace{2pt}} & \parbox[c]{4.7cm}{\vspace{2pt}$  y_1 \overline{L^c} \eta_1^* L + y_2 \overline{\psi_{2L}} \eta_3^* l_R  $\\ $+ y_3 \overline{\psi_{2L}} \eta_2^* \psi_{1R} + y_4 \overline{\psi_{1R}^c} \eta_1^* \nu_{R} $\\$+ y_e \overline{L} H l_R +\mu \eta_2 \eta_3^* \sigma^* $\vspace{2pt}} &  \color{MyGreen}{ $\cmark$} \\
\hline
T2-iv & \textbf{T2-iv-1-A} & \parbox[c]{4.7cm}{\vspace{2pt}$\eta_1 (1,1,-1,3)$\\ $\eta_2 (1,1,Y-\frac{1}{2},\alpha+3)$ \\ $\phi  (1,2,Y,\alpha)$\vspace{2pt}} &\parbox[c]{4.7cm}{\vspace{2pt}$\psi_{L,R} (1,1,1,1)$ \vspace{2pt}} & \parbox[c]{4.7cm}{\vspace{2pt}$ M_{\psi}\overline{\psi_{L}}\psi_{R}+ y_1 \overline{L^c} H^* \psi_L $\\$+ y_2 \overline{\psi_R^c} \eta_1 \nu_{R} + \lambda_1 \phi^{*} H \sigma^* \eta_2 $\\$ + \lambda_2 \phi \epsilon H \eta_1 \eta_2^* $\vspace{2pt}} &  \color{MyGreen}{ $\cmark$} \\
\hline
T2-v & \textbf{T2-v-2-A} & \parbox[c]{4.7cm}{\vspace{2pt}$\eta_1 (1,1,-1,3)$\\ $\eta_2 (1,1,Y+\frac{1}{2},\alpha-3)$\\$\phi (1,2,Y,\alpha)$\vspace{2pt}} & \parbox[c]{4.7cm}{\vspace{2pt}$\psi_{L,R} (1,1,1,1)$\vspace{2pt}} & \parbox[c]{4.7cm}{\vspace{2pt}$ M_{\psi}\overline{\psi_{L}}\psi_{R}+ y_1 \overline{L^c} H^* \psi_L $\\$+ y_2 \overline{\psi_R^c} \eta_1 \nu_{R} + \lambda_1 \phi^{*} H \eta_1 \eta_2 $\\$ + \lambda_2 \phi \epsilon H \sigma^* \eta_2^* $\vspace{2pt}} &  \color{MyGreen}{ $\cmark$}  \\
\hline
T2-vi & \textbf{T2-vi-3-A} & \parbox[c]{4.7cm}{\vspace{2pt}$\eta (1,1,-1,3)$ \vspace{2pt}}& \parbox[c]{4.7cm}{\vspace{2pt} $\psi_{1L,R} (1,1,1,1)$ \\ $\psi_{2L,R} (1,2,\frac{1}{2},-\frac{3}{2})$\\$\psi_{3L,R} (1,1,-1,\frac{3}{2})$\\$\psi_{4L,R} (1,1,0,\frac{3}{2})$\vspace{2pt}} & \parbox[c]{4.7cm}{\vspace{2pt}$ M_{\psi_1}\overline{\psi_{1L}}\psi_{1R}+ y_1 \overline{L^c} H^* \psi_{1L} $\\$+ y_2 \overline{\psi_{1R}^c} \eta \nu_{R} $\\$+ y_3 \overline{\psi_{4L/R}^c}\psi_{4L/R}\sigma^* $\\$+ y_4 \overline{\psi_{2L/R}^c}\psi_{4L/R} H^* $\\$+ y_5 \overline{\psi_{2L/R}^c}\psi_{3L/R} \epsilon H $\\$ +
y_6 \overline{\psi_{3L/R}^c}\psi_{4L/R} \eta^* $\vspace{2pt}} &  \color{MyGreen}{ $\cmark$} \\
\hline
\end{tabular}}
\end{center}\caption{ Minimal two-loop models without colored particles constructed from $d=5$ effective operator given in Eq. \eqref{d5}. Notation: the color-blind scalars $\eta$ and $\phi$ are assumed to be singlet and doublet under $SU(2)_L$.}\label{table-two}
\end{table}
%\clearpage

\FloatBarrier
\begin{figure}[th!]
\centering
$$
\includegraphics[scale=0.43]{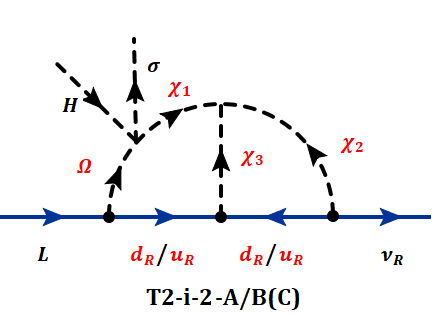}\hspace{0.1in}
\includegraphics[scale=0.43]{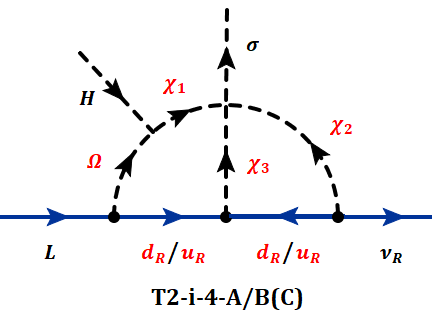}\hspace{0.1in}
\includegraphics[scale=0.43]{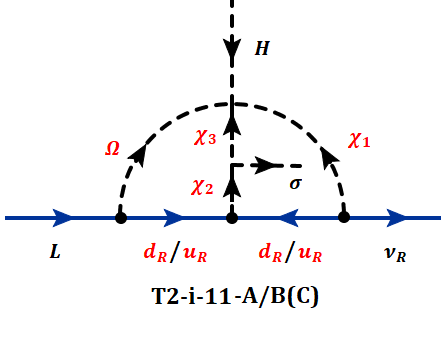}
$$
$$
\includegraphics[scale=0.43]{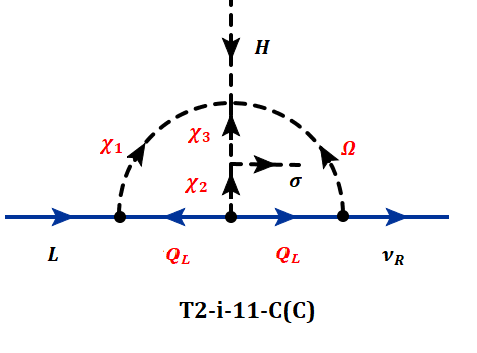}\hspace{0.1in}
\includegraphics[scale=0.43]{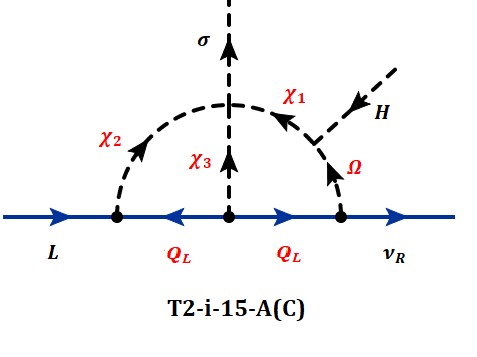}\hspace{0.1in}
\includegraphics[scale=0.41]{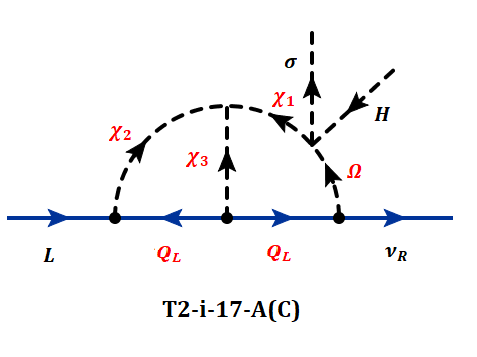}
$$
$$
\includegraphics[scale=0.43]{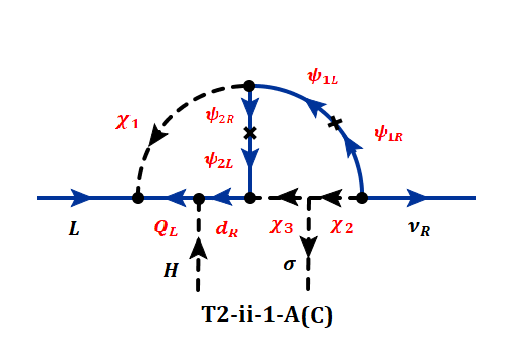}\hspace{0.1in}
\includegraphics[scale=0.43]{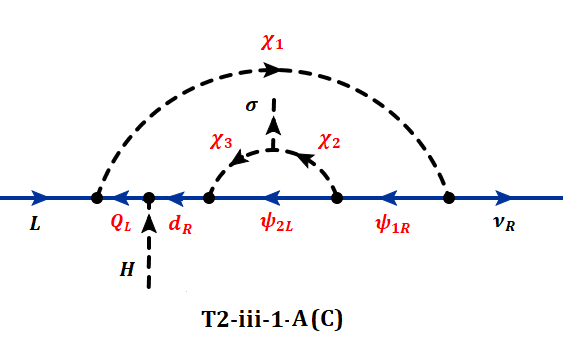}
$$
$$
\includegraphics[scale=0.43]{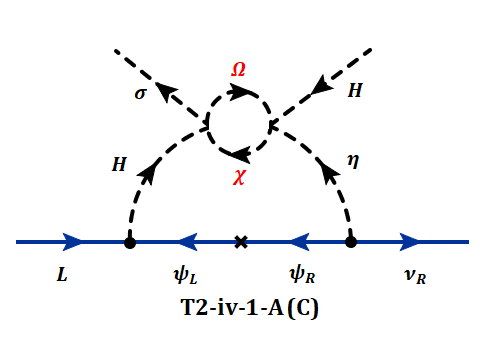}\hspace{0.1in}
\includegraphics[scale=0.43]{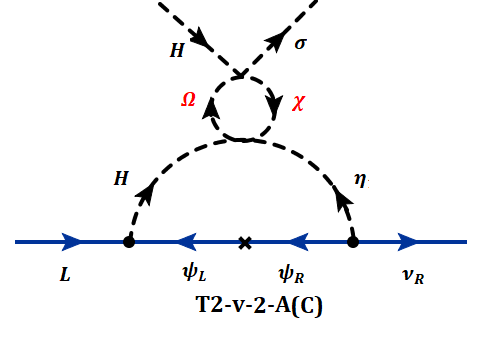}
\includegraphics[scale=0.41]{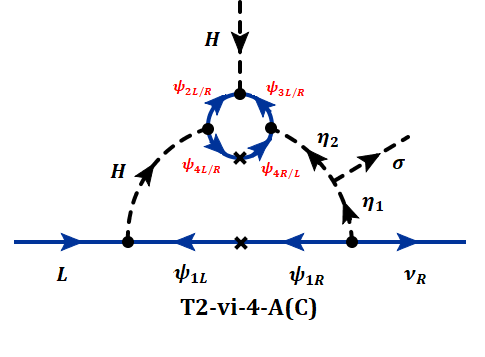}
$$
\caption{ Color version of the minimal  two-loop model diagrams arising from T-II-1,2,3,4,5 skeleton diagrams.  Quantum numbers associated with each of the model diagrams are presented in Table \ref{table-two-color}. Notation: the color-blind scalars $\eta$ and $\phi$ are assumed to be singlet and doublet under $SU(2)_L$. The  scalars,  $\chi$ and $\Omega$ charged under the color group are assumed to be $SU(2)_L$ singlet and doublet, respectively. }\label{two-f}
\end{figure}

\begin{table}[!h]
%\label{twoloop}
\begin{center}
\resizebox{1.0\textwidth}{!}{\begin{tabular}{|| c | c | c | c | c | c || }
\hline
\textbf{Skeleton} & \textbf{Models}  &\multicolumn{2}{ c |}{\textbf{New Fields}}  & \textbf{Relevant terms in Lagrangian} & \textbf{New ?}   \\ 
\cline{3-4}
 & & \textbf{Scalars} & \textbf{Fermions} &  &   \\
\hline
T2-i & \textbf{T2-i-2-A(C)} & \parbox[c]{4.7cm}{\vspace{2pt} $\textcolor{red}{\chi_1} (\bar{3},1,\frac{1}{3},-\frac{13}{3})$\\
 $\textcolor{red}{\chi_2} (\bar{3},1,\frac{1}{3},\frac{11}{3})$\\
 $ \textcolor{red}{\chi_3} (\bar{3},1,-\frac{2}{3},\frac{2}{3})$ \\
 $\textcolor{red}{\Omega}  (\bar{3},2,-\frac{1}{6},-\frac{4}{3})$\vspace{2pt}} &--- &
 \parbox[c]{4.7cm}{\vspace{2pt} $y_1\overline{L} \Omega d_R 
 + y_2 \overline{d^c_R} \chi_3^* d_R $\\ $+ y_3 \overline{d^c_R} \chi_2 \nu_R + \mu \chi_1 \chi_2 \chi_3 $\\$+ \lambda \Omega \epsilon H \sigma^* \chi_1^*  $\vspace{2pt}} &  \color{MyGreen}{ $\cmark$} \\
  \cline{2-6}
  & \textbf{T2-i-2-B(C)} &  \parbox[c]{4.7cm}{\vspace{2pt}$\textcolor{red}{\chi_1} (\bar{3},1,-\frac{2}{3},-\frac{13}{3})$\\
  $\textcolor{red}{\chi_2} (\bar{3},1,-\frac{2}{3},\frac{11}{3})$\\
  $\textcolor{red}{\chi_3} (\bar{3},1,\frac{4}{3},\frac{2}{3})$ \\
 $\textcolor{red}{\Omega}  (\bar{3},2,-\frac{7}{6},-\frac{4}{3})$\vspace{2pt}} &--- & 
  \parbox[c]{4.7cm}{\vspace{2pt} $y_1\overline{L} \Omega u_R 
 + y_2 \overline{u^c_R} \chi_3^* u_R $\\ $+ y_3 \overline{u^c_R} \chi_2 \nu_R + \mu \chi_1 \chi_2 \chi_3 $\\$+ \lambda \Omega \epsilon H \sigma^* \chi_1^*  $\vspace{2pt}} &  \color{MyGreen}{ $\cmark$} \\
  \cline{2-6}
  &  \textbf{T2-i-4-A(C)} &  \parbox[c]{4.7cm}{\vspace{2pt}$\textcolor{red}{\chi_1} (\bar{3},1,\frac{1}{3},-\frac{4}{3})$ \\
  $\textcolor{red}{\chi_2} (\bar{3},1,\frac{1}{3},\frac{11}{3})$ \\
  $\textcolor{red}{\chi_3} (\bar{3},1,-\frac{2}{3},\frac{2}{3})$\\
  $\textcolor{red}{\Omega}  (\bar{3},2,-\frac{1}{6},-\frac{4}{3})$\vspace{2pt}} &--- & 
  \parbox[c]{4.7cm}{\vspace{2pt}$y_1 \overline{L} \Omega d_R 
+ y_2 \overline{d^c_R} \chi_3^* d_R $\\ $+ y_3 \overline{d^c_R} \chi_2 \nu_R + \mu \Omega \epsilon H \chi_1^* $\\$+ \lambda \chi_1 \chi_2 \chi_3 \sigma^* $\vspace{2pt}} &  \color{MyGreen}{ $\cmark$} \\
 \cline{2-6}
  &  \textbf{T2-i-4-B(C)} &  \parbox[c]{4.7cm}{\vspace{2pt} $\textcolor{red}{\chi_1} (\bar{3},1,-\frac{2}{3},-\frac{4}{3})$ \\
 $\textcolor{red}{\chi_2} (\bar{3},1,-\frac{2}{3},\frac{11}{3})$ \\
 $\textcolor{red}{\chi_3} (\bar{3},1,\frac{4}{3},\frac{2}{3})$ \\
 $\textcolor{red}{\Omega}  (\bar{3},2,-\frac{7}{6},-\frac{4}{3})$\vspace{2pt}} &--- &
 \parbox[c]{4.7cm}{\vspace{2pt}$y_1 \overline{L} \Omega u_R 
+ y_2 \overline{u^c_R} \chi_3^* u_R $\\ $+ y_3 \overline{u^c_R} \chi_2 \nu_R + \mu \Omega \epsilon H \chi_1^* $\\$+ \lambda \chi_1 \chi_2 \chi_3 \sigma^* $\vspace{2pt}} &  \color{MyGreen}{ $\cmark$} \\
 \cline{2-6}
  & \textbf{T2-i-11-A(C)} & \parbox[c]{4.7cm}{\vspace{2pt}$\textcolor{red}{\chi_1} (\bar{3},1,\frac{1}{3},\frac{11}{3})$ \\
 $\textcolor{red}{\chi_2} (\bar{3},1,-\frac{2}{3},\frac{2}{3})$  \\
 $\textcolor{red}{\chi_3} (\bar{3},1,-\frac{2}{3},-\frac{7}{3})$ \\
 $\textcolor{red}{\Omega}  (\bar{3},2,-\frac{1}{6},-\frac{4}{3})$ \vspace{2pt}} &--- & 
 \parbox[c]{4.7cm}{\vspace{2pt}$y_1 \overline{L} \Omega d_R 
 + y_2 \overline{d^c_R} \chi_2^* d_R $\\ $+ y_3 \overline{d^c_R} \chi_1 \nu_R + \mu \chi_2 \chi_3^* \sigma^* $\\$+ \lambda \Omega \epsilon H \chi_1 \chi_3 $\vspace{2pt}}&  \color{MyGreen}{ $\cmark$}  \\
  \cline{2-6}
  & \textbf{T2-i-11-B(C)} & \parbox[c]{4.7cm}{\vspace{2pt}$\textcolor{red}{\chi_1} (\bar{3},1,-\frac{2}{3},\frac{11}{3})$ \\
 $\textcolor{red}{\chi_2} (\bar{3},1,\frac{4}{3},\frac{2}{3})$  \\
 $\textcolor{red}{\chi_3} (\bar{3},1,\frac{4}{3},-\frac{7}{3})$ \\
 $\textcolor{red}{\Omega}  (\bar{3},2,-\frac{7}{6},-\frac{4}{3})$  \vspace{2pt}} &--- & 
 \parbox[c]{4.7cm}{\vspace{2pt}$y_1 \overline{L} \Omega u_R 
 + y_2 \overline{u^c_R} \chi_2^* u_R $\\ $+ y_3 \overline{u^c_R} \chi_1 \nu_R + \mu \chi_2 \chi_3^* \sigma^* $\\$+ \lambda \Omega \epsilon H \chi_1 \chi_3 $\vspace{2pt}} &  \color{MyGreen}{ $\cmark$} \\
  \cline{2-6}
  & \textbf{T2-i-11-C(C)} & \parbox[c]{4.7cm}{\vspace{2pt}$\textcolor{red}{\chi_1} (3,1,-\frac{1}{3},-\frac{2}{3})$ \\
 $\textcolor{red}{\chi_2} (3,1,-\frac{1}{3},-\frac{2}{3})$  \\
 $\textcolor{red}{\chi_3} (3,1,-\frac{1}{3},-\frac{11}{3})$ \\
 $\textcolor{red}{\Omega}  (3,2,\frac{1}{6},\frac{13}{3})$ \vspace{2pt}} &--- &
 \parbox[c]{4.7cm}{\vspace{2pt}$y_1 \overline{L^c} \epsilon \chi_1^* Q_L 
 + y_2 \overline{Q^c_L} \epsilon Q_L \chi_2$\\ $+ y_3 \overline{Q_L} \Omega \nu_R + \mu \chi_2 \chi_3^* \sigma^* $\\$+ \lambda \Omega \epsilon H \chi_1 \chi_3 $\vspace{2pt}} & \color{MyGreen}{ $\cmark$} \\
 \cline{2-6}
  & \textbf{T2-i-15-A(C)} & \parbox[c]{4.7cm}{\vspace{2pt}$\textcolor{red}{\chi_1} (3,1,\frac{2}{3},\frac{13}{3})$  \\
 $\textcolor{red}{\chi_2} (3,1,-\frac{1}{3},-\frac{2}{3})$ \\
 $\textcolor{red}{\chi_3} (3,1,-\frac{1}{3},-\frac{2}{3})$  \\
 $\textcolor{red}{\Omega}  (3,2,\frac{1}{6},\frac{13}{3})$ \vspace{2pt}} &--- &
 \parbox[c]{4.7cm}{\vspace{2pt}$y_1 \overline{L^c} \epsilon \chi_2^* Q_L 
 + y_2 \overline{Q^c_L} \epsilon Q_L \chi_3 $\\ $+ y_3 \overline{Q_L} \Omega \nu_R + \mu  \chi_1^* \Omega \epsilon H  $\\$+ \lambda \chi_1 \chi_2 \chi_3 \sigma^*  $\vspace{2pt}} &  \color{MyGreen}{ $\cmark$} \\
 \cline{2-6}
  & \textbf{T2-i-17-A(C)} & \parbox[c]{4.7cm}{\vspace{2pt}$\textcolor{red}{\chi_1} (3,1,\frac{2}{3},\frac{4}{3})$  \\
 $\textcolor{red}{\chi_2} (3,1,-\frac{1}{3},-\frac{2}{3})$  \\
 $\textcolor{red}{\chi_3} (3,1,-\frac{1}{3},-\frac{2}{3})$ \\
 $\textcolor{red}{\Omega}  (3,2,\frac{1}{6},\frac{13}{3})$ \vspace{2pt}} &--- &
 \parbox[c]{4.7cm}{\vspace{2pt}$y_1 \overline{L^c} \epsilon \chi_2^* Q_L 
 + y_2 \overline{Q^c_L} \epsilon Q_L \chi_3 $\\ $+ y_3 \overline{Q_L} \Omega \nu_R + \mu \chi_1 \chi_2 \chi_3 $\\$+ \lambda \sigma^* \chi_1^* \Omega \epsilon H   $\vspace{2pt}} &  \color{MyGreen}{ $\cmark$} \\
\hline
T2-ii & \textbf{T2-ii-1-A(C)} &  \parbox[c]{4.7cm}{\vspace{2pt} $\textcolor{red}{\chi_1}  (\bar{3},1,\frac{1}{3},\frac{2}{3})$\\$\textcolor{red}{\chi_2} (\bar{3},1,-Y-\frac{1}{3},\frac{10}{3}-\alpha)$\\    $\textcolor{red}{\chi_3} (\bar{3},1,-Y-\frac{1}{3},\frac{1}{3}-\alpha)$\vspace{2pt}} & 
 \parbox[c]{4.7cm}{\vspace{2pt} $\textcolor{red}{\psi_{1L,R}} (3,1,Y+\frac{1}{3},\alpha+\frac{2}{3})$\\ $\textcolor{red}{\psi_{2L,R}} (\bar{3},1,Y,\alpha)$ \vspace{2pt}}  &
 \parbox[c]{4.7cm}{\vspace{2pt}$ M_{\psi_1}\overline{\psi_{1L}}\psi_{1R} +M_{\psi_2}\overline{\psi_{2L}}\psi_{2R}$\\ $+ y_1 \overline{L^c} \epsilon \chi_1 Q_L 
+ y_2 \overline{\psi_{2L}} \chi_3^* d_R $\\ $+ y_3 \overline{\psi^c_{1R}} \chi_2 \nu_R + y_4 \overline{\psi_{1L}} \chi_1 \psi_{2R} $\\$+ y_d \overline{Q_L} H d_R +\mu \chi_2 \chi_3^* \sigma^* $\vspace{2pt}} &  \color{MyGreen}{ $\cmark$} \\
\hline
T2-iii & \textbf{T2-iii-1-A(C)} & \parbox[c]{4.7cm}{\vspace{2pt} $\textcolor{red}{\chi_1} (3,1,-\frac{1}{3},-\frac{2}{3})$\\ $\textcolor{red}{\chi_2} (\bar{3},1,-(Y+\frac{1}{3}),\frac{10}{3}-\alpha)$\\$\textcolor{red}{\chi_3} (\bar{3},1,-(Y+\frac{1}{3}),\frac{1}{3}-\alpha)$ \vspace{2pt}} &
 \parbox[c]{4.7cm}{\vspace{2pt}$ \textcolor{red}{\psi_{1L,R}} (3,1,-\frac{1}{3},\frac{10}{3})$\\ $\textcolor{red}{\psi_{2L,R}}(\bar{3},1,Y,\alpha)$ \vspace{2pt}}  & \parbox[c]{4.7cm}{\vspace{2pt}$ y_1 \overline{Q_L^c} \chi_1^* L + y_2 \overline{\psi_{2L}} \chi_3^* d_R $\\ $+ y_3 \overline{\psi_{2L}} \chi_2^* \psi_{1R} + y_4 \overline{\psi_{1R}^c} \chi_1^* \nu_{R} $\\$ + y_d \overline{Q_L} H d_R + \mu \chi_2 \chi_3^* \sigma^* $\vspace{2pt}} &  \color{MyGreen}{ $\cmark$} \\
\hline
T2-iv & \textbf{T2-iv-1-A(C)} & \parbox[c]{4.7cm}{\vspace{2pt}$\eta (1,1,-1,3)$\\$\textcolor{red}{\chi} (3,1,Y-\frac{1}{2},\alpha+3)$\\$\textcolor{red}{\Omega}  (3,2,Y,\alpha)$\vspace{2pt}} & \parbox[c]{4.7cm}{\vspace{2pt}$\psi_{L,R} (1,1,1,1)$\vspace{2pt}} &
 \parbox[c]{4.7cm}{\vspace{2pt}$ M_{\psi}\overline{\psi_{L}}\psi_{R}+ y_1 \overline{L^c} H^* \psi_L $\\$+ y_2 \overline{\psi_R^c} \eta \nu_{R} + \lambda_1 \Omega^{*} H \sigma^* \chi $\\$ + \lambda_2 \Omega \epsilon H \eta \chi^* $\vspace{2pt}} &  \color{MyGreen}{ $\cmark$} \\
\hline
T2-v & \textbf{T2-v-2-A(C)} & \parbox[c]{4.7cm}{\vspace{2pt}$\eta_1 (1,1,-1,3)$\\$\textcolor{red}{\chi} (3,1,Y+\frac{1}{2},\alpha-3)$\\$\textcolor{red}{\Omega}(3,2,Y,\alpha)$\vspace{2pt}} & \parbox[c]{4.7cm}{\vspace{2pt}$\psi_{L,R} (1,1,1,1)$\vspace{2pt}} & 
 \parbox[c]{4.7cm}{\vspace{2pt}$ M_{\psi}\overline{\psi_{L}}\psi_{R}+ y_1 \overline{L^c} H^* \psi_L $\\$+ y_2 \overline{\psi_R^c} \eta \nu_{R} + \lambda_1 \Omega^{*} H \eta \chi $\\$ + \lambda_2 \Omega \epsilon H \sigma^* \chi^* $\vspace{2pt}} &  \color{MyGreen}{ $\cmark$} \\
\hline
T2-vi & \textbf{T2-vi-4-A(C)} & \parbox[c]{4.7cm}{\vspace{2pt}$\eta_1 (1,1,-1,3)$\\ $\eta_2 (1,1,-1,0)$ \vspace{2pt}} & \parbox[c]{4.7cm}{\vspace{2pt}$\psi_{1L,R} (1,1,1,1)$\\ $\textcolor{red}{\psi_{2L,R}} (3,2,Y,\alpha)$\\$\textcolor{red}{\psi_{3L,R}} (\bar{3},1,-Y-\frac{1}{2},-\alpha)$\\ $\textcolor{red}{\psi_{4L,R}} (\bar{3},1,-Y+\frac{1}{2},-\alpha)$\vspace{2pt}} &
\parbox[c]{4.7cm}{\vspace{2pt}$ M_{\psi_1}\overline{\psi_{1L}}\psi_{1R}+M_{\psi_4}\overline{\psi_{4L}}\psi_{4R} $\\$ + y_1 \overline{L^c} H^* \psi_{1L} + y_2 \overline{\psi_{1R}^c} \eta_1 \nu_{R} $\\$+ y_3 \overline{\psi_{2L/R}^c}\psi_{4L/R}H^* $\\$+ y_4 \overline{\psi_{2L/R}^c}\psi_{3L/R}\epsilon H $\\$+ y_5 \overline{\psi_{4R/L}}\psi_{3L/R} \eta_2^* +
\mu \eta_1 \eta_2^* \sigma^* $\vspace{2pt}} &  \color{MyGreen}{ $\cmark$} \\
\hline
    
\end{tabular}}
\end{center}\caption{ Minimal two-loop models with colored particles constructed from $d=5$ effective operator given in Eq. \eqref{d5}. Each of the models contains $\sigma (1,1,0,3)$ that breaks the $U(1)_{B-L}$ symmetry. Notation: the color-blind scalars $\eta$ and $\phi$ are assumed to be singlet and doublet under $SU(2)_L$. The  scalars,  $\chi$ and $\Omega$ charged under the color group are assumed to be $SU(2)_L$ singlet and doublet, respectively.  }\label{table-two-color} 
\end{table}
\clearpage

%%%%%%%%%%%%%%%%%%%%%%%%%%%%%%%%%%%%%%%%%% 
%%%%%%%%%%%%%%%%%%%%%%%%%%%%%%%%%%%%%%%%%% 
%%%%%%%%%%%%%%%%%%%%%%%%%%%%%%%%%%%%%%%%%% 
\subsection{Search for minimal  two-loop models with colored particles} 
In this section, we carry out our search for finding  minimal two-loop models from  various topologies introduced above by employing colored particles in the internal lines. 

\begin{itemize}
    \item[$\square$] \underline{T2-i-y-A/B/C(C)}:
\end{itemize}\vspace{-10pt}

We start with the T2-i skeleton diagram as shown in Fig. \ref{two-b}. As already explained above, this leads to 43 diagrams T2-i-y (y=1-43) presented in Fig. \ref{two-c}, out of which, we have identified  3 minimal candidate models  for the scenario with color singlet states. The case with colored particles running in the loops is somewhat different and without introducing any new fermions, we find 9 candidates. Two such cases are built out of T2-i-2 diagram, where the internal fermions are either $d_R$ or $u_R$ (which are labelled as T2-i-A(C) and T2-i-2-B(C)). The  same thing can be repeated for both T2-i-4 and T2-i-11, resulting in another 4 more minimal models (labelled as T2-i-4-A/B(C) and T2-i-11-A/B(C)). On the contrary, when the internal fermion line is taken to be $Q_L$ instead of either $d_R$ or $u_R$, in addition to T2-i-11, two more different models can be fabricated out of T2-i-15 and T2-i-17 diagrams that also do not require any fermions BSM. All these 9 different models are presented in Fig. \ref{two-f} and their particle contents are summarized in Table \ref{table-two}. Note that in these models, the scalars carrying colors are triplets (or anti-triplets), which is in fact, the lowest dimensional representation under $SU(3)_C$. 

One can immediately construct variations of these models by replacing the color anti-triplet $\chi_3$ in T2-i-y-z(C) (y=2,4,11, z=A,B)  and  the color anti-triplet $\chi_2$ in T2-i-y-z(C) (y=11, z=A,B) by  $SU(3)_C$ sextets \textcolor{black}{(since for $SU(3)$ group,  $\overline{3}\times \overline{3}\supset \overline{6}$ and  $6\times \overline{6}\supset 1$)}, which however are not listed in Table \ref{table-two-color} due to our minimality arguments. Similar variation can be done for T2-i-11-C(C) by replacing the color triplets $\chi_{2,3}$ by color sextets and in T2-i-y-z(C) (with y=15,17 and z=A) by replacing the color triplet $\chi_{3}$ by a color sextet. 

\begin{itemize}
    \item[$\square$] \underline{T2-ii-1-A(C) $\&$ T2-iii-1-A(C)}:
\end{itemize}\vspace{-10pt}

Colored extension of models arising from T2-ii and T2-iii skeleton diagrams can be done straightforwardly following the discussions presented above for models with color singlet states. To do so,   $L$ and $\ell_R$ in T2-ii-1-A are replaced by $Q_L$ and $d_R$ that results  in T2-ii-1-A(C) model diagram. Subsequently, the rest of the internal particles must carry color (color triplets or anti-triplets). Similarly, in T2-iii-1-A, $L\leftrightarrow Q_L$ and $\ell_R\leftrightarrow d_R$ replacements are made in the internal fermion line  adjacent to the incoming $L$ line. As already pointed out, due to the chosen SM Higgs insertion\footnote{It is noteworthy to mention that attaching $H$ at the outer scalar loop would require a colored triplet scalar which is an iso-doublet. By inserting $H$ in this manner, two separate next-to-minimal models can be constructed depending on whether $d_R$ or $u_R$ is running in the internal fermion propagator adjacent to incoming $L$. However, we do not list these scenarios since they do not qualify to be the most minimal models.}, this model does not require any BSM scalar doublet. Completion of this model diagram is made by utilizing three BSM colored triplet scalars and two colored triplet fermions, which are all iso-singlets. See Fig. \ref{two-f} and Table \ref{table-two-color} for more details. 

Note that if the color anti-triplets $\psi_1, \chi_{2,3}$ are replaced by color triplets in T2-ii-1-A(C), then a variation of this model can be constructed with $\psi_2$ being octet under $SU(3)_C$. A   similar variant model can emerge from  T2-iii-1-A(C) if $\chi_{2,3}$ be the color triplets instead and $\psi_2$ being color octet under $SU(3)_C$.  

\begin{itemize}
    \item[$\square$] \underline{T2-iv-1-A(C) $\&$ T2-v-2-A(C)}:
\end{itemize}\vspace{-10pt}

The fabrication process of T2-iv and T2-v are discussed at length in the previous section. It is pointed out that two of the internal scalar propagators connected to the fermions must be $H$ and $\eta$, no other variation can be made to construct a viable two-loop model out of these skeleton diagrams. Due to this special property, the only two scalars that can carry color quantum numbers are $\chi$ and $\omega$, as shown in Fig. \ref{two-f}. To satisfy the minimality conditions, these fields are chosen to be color triplets, see Table \ref{table-two-color} for the complete quantum numbers of all the states associated to these models. 

\begin{itemize}
    \item[$\square$] \underline{T2-vi-4-A(C)}:
\end{itemize}  

Note that, for the scenario with color singlets, the minimal model is constructed out of T2-vi-3 diagram (Fig. \ref{two-h}), whereas, if particles carrying colors are allowed, no such vertex of the type $\psi\psi\sigma$ is allowed, since it cannot be invariant under $SU(3)_C$. This can only be done if two different fermions are used to form this vertex, hence fails to be minimal. Instead, we build the minimal model by making use of T2-vi-4 diagram (instead of T2-vi-3, see Fig. \ref{two-h}), which requires two singly charged scalars (instead of one), but the same number of BSM fermions as before.  For minimality, three of the fermions that are allowed to carry color charge are taken to be  triplets (or anti-triplets) under $SU(3)_C$. See Table \ref{table-two-color} for more details.

%%%%%%%%%%%%%%%%%%%%%%%%%%%%%%%%%%%%%%%%%%%%%%%
\section{Possible dark matter candidates}\label{SEC-dark}
To generate neutrino mass, new particles must be added to the theory and we have presented many such examples in the previous two sections. From the theoretical point of view, it is very appealing, if the added particles needed to generate neutrino mass can also serve as DM candidates. Though our main focus of this work is to select minimal models of Dirac neutrino mass arising from generic one-loop and two-loop topologies constructed from the $d=5$ effective operator of Eq. \eqref{d5}, in this section, we investigate  the possibilities of having DM particles within these minimal scenarios. Since a DM particle cannot carry color, so we only concentrate on the models, where color singlet states are employed to generate neutrino mass.  

In radiative neutrino mass models, DM particles can be incorporated, if additional symmetries beyond the SM are imposed. This idea was first proposed in Ref. \cite{Ma:2006km} for Majorana neutrinos and known as \textit{scotogenic} model. It was shown that a simple application of $\mathcal{Z}_2$ symmetry does the job, as long as, only BSM particles are circulating inside the loop.  This  $\mathcal{Z}_2$ symmetry must remain unbroken even after the breaking of EW symmetry.   

In the framework we are working, the spontaneously broken $U(1)_{B-L}$ symmetry, may leave a residual unbroken symmetry that can potentially stabilize the DM particle. The nature of the unbroken residual symmetry $U(1)_{B-L}\to \mathcal{Z}_N$ depends on the details of the $B-L$ charge assignments of the scalars. In our set-up, since, the SM Higgs doublet $H$ is neutral under  $U(1)_{B-L}$, the nature of the unbroken symmetry is entirely determined by the charge of the singlet scalar $\sigma$. In scenarios, where all the particles carry integer $B-L$ charge, then $U(1)_{B-L}\to \mathcal{Z}_3$ is realized because, $\sigma$ carries three units of $B-L$ charge. Models with $N=3$ cannot stabilize the DM particle, if any, by only the residual symmetry. \textcolor{black}{In fact, it has been  shown \cite{Bonilla:2018ynb, Bonilla:2019hfb}  that in scotogenic Dirac neutrino mass models,  for  any odd integer $N$, one can always write down  operators that eventually lead to DM decays.}        We find that three out of eight one-loop models (see Table \ref{DM-i}) and three out of nine two-loop models (see Table \ref{DM-ii})  have the $\mathcal{Z}_3$ residual lepton symmetry, and consequently,  DM stability cannot be guaranteed by the residual symmetry alone. 

On the contrary, in models where any of the particles is carrying a half-odd-integer $B-L$ charge, the rescaled charge of $\sigma$ must be 6, hence the unbroken symmetry is  $U(1)_{B-L}\to \mathcal{Z}_6$ instead of $\mathcal{Z}_3$. In these models, this unbroken  $\mathcal{Z}_6$ plays the role of residual dark symmetry, hence stabilizes the DM candidates. Most of the models presented in this work, permit such a choice of half-odd-integer $B-L$ charge and, consequently, naturally incorporates DM within the minimal set-up without the need of extending the particle content any further. It is interesting to note that in the two-loop models that allow the   half-odd-integer $B-L$ charge assignment, DM candidate exists, even though SM fermions are propagating inside the loop, a feature that has been overlooked in the literature. List of the models and their associated residual symmetries and possible DM candidates are lsited in Tables \ref{DM-i} and  \ref{DM-ii}. 
In a future work \cite{future}, we plan to construct the minimal \textit{scotogenic} one-loop and two-loop models of Dirac neutrino mass arising from this working framework and study their associated DM phenomenology in great details.

%%%%%%%%%%%%%%%%%%%%%%%%%%%%%%%%%%%%%%%%%%%%%%%%%%%%%%
\FloatBarrier
\begin{table}[t!] 
\centering
\footnotesize
\resizebox{0.9\textwidth}{!}{
\begin{tabular}{|c|c|c|c|c|}
\hline 
Models&  \pbox{10cm}{
\vspace{2pt}
Residual lepton \\symmetry
\vspace{3pt}
} &\pbox{10cm}{
\vspace{2pt}
Residual dark \\symmetry
\vspace{3pt}
} 
&\pbox{10cm}{
\vspace{2pt}
Choice of $Y$
\vspace{3pt}
}
&\pbox{10cm}{
\vspace{2pt}
Possible DM \\candidate
\vspace{3pt}
}
\\ \hline\hline
T1-i-1-A&$\mathcal{Z}_3$&\color{red}{ $\xmark$}&---&\color{red}{ $\xmark$}\\ \hline
T1-i-1-B&$\mathcal{Z}_3$&\color{red}{ $\xmark$}&---&\color{red}{ $\xmark$}\\ \hline
T1-i-1-C&$\mathcal{Z}_6$&\color{MyGreen}{ $\cmark$}&---&$\psi_{L,R}, \eta, \phi$\\ \hline
T1-i-2-A&$\mathcal{Z}_6$&\color{MyGreen}{ $\cmark$}&
$0$&$\psi_{L,R}, \phi, \eta_{1},\eta_{2}$ \\ 
\cline{4-5} 
&&&$-1$&$\phi$\\ \hline
T1-i-2-B&$\mathcal{Z}_6$&\color{MyGreen}{ $\cmark$}&
$0$&$\psi_{L,R}, \phi_{1},\phi_{2},\eta_{1},\eta_{2}$ \\
\cline{4-5} 
&&&$-1$&$\phi_{1},\phi_{2}$\\ \hline
T1-i-3-A&$\mathcal{Z}_3$&\color{red}{ $\xmark$}&---&\color{red}{ $\xmark$}\\ \hline
T1-i-3-B&$\mathcal{Z}_6$&\color{MyGreen}{ $\cmark$}&---&$\psi_{1L,R},\psi_{2L,R},\phi$ \\ 
 \hline
T1-ii-1-A&$\mathcal{Z}_6$&\color{MyGreen}{ $\cmark$}&
$0$&$\psi_{L,R}, \eta, \phi$\\ 
\cline{4-5} 
&&&$-1$&$\phi$\\ \hline
\end{tabular}
}
\caption{ As an example, in this table, for  models with undetermined $\alpha$, we take $\alpha=n/2$, where $n$ is an odd integer.    
}\label{DM-i}
\end{table}
%%%%%%%%%%%%%%%%%%%%%%%%%%%%%%%%%%%%%%%%%%%%%%%%%%%%%

%%%%%%%%%%%%%%%%%%%%%%%%%%%%%%%%%%%%%%%
%%%%%%%%%%%%%%%%%%%%%%%%%%%%%%%%%%%%%%%
%%%%%%%%%%%%%%%%%%%%%%%%%%%%%%%%%%%%%%%%%%%%%%%%%%%%%%
\FloatBarrier
\begin{table}[t!] 
\centering
\footnotesize
\resizebox{0.9\textwidth}{!}{
\begin{tabular}{|c|c|c|c|c|}
\hline
Models&  \pbox{10cm}{
\vspace{2pt}
Residual lepton \\symmetry
\vspace{3pt}
} &\pbox{10cm}{
\vspace{2pt}
Residual dark \\symmetry
\vspace{3pt}
} 
&\pbox{10cm}{
\vspace{2pt}
Choice of $Y$
\vspace{3pt}
}
&\pbox{10cm}{
\vspace{2pt}
Possible DM \\candidate
\vspace{3pt}
}
\\ \hline\hline
T2-i-2-A&$\mathcal{Z}_3$&\color{red}{ $\xmark$}&---&\color{red}{ $\xmark$}\\ \hline
T2-i-4-A&$\mathcal{Z}_3$&\color{red}{ $\xmark$}&---&\color{red}{ $\xmark$}\\ \hline
T2-i-11-A&$\mathcal{Z}_3$&\color{red}{ $\xmark$}&---&\color{red}{ $\xmark$}\\ \hline
T2-ii-1-A&$\mathcal{Z}_6$&\color{MyGreen}{ $\cmark$}&---&$\psi_{1L,R},\eta_2$ \\ 
\hline
T2-ii-2-A&$\mathcal{Z}_6$&\color{MyGreen}{ $\cmark$}&---&$\psi_{2L,R}$ \\ 
\hline
T2-iii-1-A
&$\mathcal{Z}_6$&\color{MyGreen}{ $\cmark$}&
$0$&$\psi_{2L,R}$ \\ 
\cline{4-5} 
&&&$-1$&$ \eta_{2},\eta_{3}$\\
\hline
T2-iv-1-A
&$\mathcal{Z}_6$&\color{MyGreen}{ $\cmark$}&
$\frac{1}{2}$&$\phi, \eta$ \\
\cline{4-5} 
&&&$-\frac{1}{2}$&$\phi$\\
\hline
T2-v-2-A
&$\mathcal{Z}_6$&\color{MyGreen}{ $\cmark$}&
$\frac{1}{2}$&$\phi$ \\ 
\cline{4-5} 
&&&$-\frac{1}{2}$&$\phi, \eta_2$\\
\hline
T2-vi-3-A
&$\mathcal{Z}_6$&\color{MyGreen}{ $\cmark$}&
---&$\psi_{2L,R},\psi_{4L,R}$ \\
\hline
\end{tabular}
}
\caption{ As an example, in this table, for  models with undetermined $\alpha$, we take $\alpha=n/2$, where $n$ is an odd integer. 
}\label{DM-ii}
\end{table}

%%%%%%%%%%%%%%%%%%%%%%%%%%%%%%%%%%%%%%%%%%%%%%%
\section{Conclusions}\label{SEC-con}
In this work,  
we have performed a systematic search for  the minimal Dirac neutrino mass models that arise from generic one-loop and two-loop topologies 
constrcuted from the $d=5$ effective operator: $\overline{L} \widetilde{H}\;{\nu_R}\sigma$. Dirac nature of neutrinos is ensured by the $U(1)_{B-L}$ gauge extension of SM and 
with non-universal charge assignments of the right-handed neutrinos under it. After setting up our ground rules for minimality within the working framework, we have done an exhaustive analysis, case by case, to select economical models with and without colored particles propagating in the loop(s). 

At one-loop level, out of the six possible topologies T1-x (x=i-vi), we have found only two topologies T1-i and T1-ii that can lead to viable  minimal models for Dirac neutrinos. After filtering out the less economical models, we ended up finding  seven minimal models from T1-i and a single model from T1-ii with color singlet states. The corresponding numbers are eight and one, when colored particles are employed to build  models.  
The complete list of viable one-loop models are  listed in Tables \ref{table-one} and \ref{table-one-color},  and the representative neutrino mass generation diagrams are presented in Figs. \ref{one-d} and \ref{one-e}.

In the two-loop scenario,  there are essentially eighteen  topologies that can generate a potentially very large number of diagrams. We thoroughly  investigate the six basic diagrams T2-x (x=i-vi) that emerge from the generic topologies,  and named them skeleton diagrams. As discussed  in the text, many variations of models can be constructed from  each of these  skeleton diagrams. Our systematic procedure, discussed at great length,  helped us selecting the associated minimal versions of the models. We have found  three minimal models from T2-i, two from T2-ii and one each from T2-x with x=ii-vi skeleton diagrams by employing particles that do not carry color.  When the internal particles are assumed to be non-singlets of color group, we have found   nine minimal models from T2-i and one from each T2-x with x=ii - vi skeleton diagrams.    The complete list of two-loop models is summarized in Tables \ref{table-two} and \ref{table-two-color}, and the representative neutrino mass generation diagrams are shown in Figs. \ref{two-e} and  \ref{two-f}. 

Furthermore, we have explored the possibilities of having  DM candidates in each of these models.   In the one-loop models, where SM particles circulate in the loop, there is no prospect of having a DM particle. In the alternate scenario, where all BSM states are running inside the loop, appearance of DM candidate arise naturally. Out of the eight models without colored states, five of them can incorporate DM particles. Whereas for the two-loop models, six out of nine minimal models contain DM candidate. It is interesting to note that in the two-loop scenarios, even though, SM particles circulate in the loops, DM candidate can still be obtained as a result of  emergent dark symmetry. The models with colored states do not lead to DM particles since, stable colored particles are experimentally excluded. In conclusion, dark matter candidate, naturally arises in most of the minimal models presented here, without extending the particle content any further.  In our set-up, the stability of these DM particles are guaranteed by the residual dark symmetry emerging from the spontaneously breaking of the $U(1)_{B-L}$ symmetry.

In total, we have worked out  forty unique minimal models.  
Whereas only a few of these models presented in this work are discussed in the literature before, most of the models are new.  It is interesting to note that every single minimal model presented in this work with colored particles only require fundamental representation under $SU(3)_C$. Furthermore, for all models with or without colored particles, no representation higher than a fundamental representation  of $SU(2)_L$ are employed either.  Each of the models presented in this work can have very distinct and rich phenomenology and must be studied case by case,  which is beyond the scope of this work. Following our detailed systematic approach presented in this work, new models of radiative Dirac neutrino mass can be constructed  straightforwardly in  different frameworks, such as by implementing alternative symmetries (discrete or continuous, for example $U(1)_R$ or general $U(1)_X$ theories) to forbid tree-level Dirac and  Majorana mass terms at all orders.

%%%%%%%%%%%%%%%%%%%%%%%%%%%%%%%%%%%%%%%%%%%%%%%%%%%%%
%%%%%%%%%%%%%%%%%%%%%%%%%%%%%%%%%%%%%%%%%%%%%
\section*{Acknowledgments}
We are grateful for helpful discussions with Kaladi Babu and Ernest Ma. This work was partially supported by  the US Department of Energy Grant Number DE-SC 0016013 and also by the Neutrino Theory Network Program under Grant No. DE-AC02-07CH11359. SJ thanks the Fermilab Theory Group for warm hospitality during the completion of this work.

%%%%%%%%%%%%%%%%%%%%%%%%%%%
\bibliographystyle{utphys}
\bibliography{reference}

\end{document}